\documentclass[twocolumn,prc,floatfix,superscriptaddress,longbibliography]{revtex4-1}
\usepackage[dvips]{graphicx}
\usepackage{bm,color}
\usepackage{dcolumn}

\def\lamb#1#2{$^{#1}_{\Lambda}${#2}} 
\def\lam#1#2{$^{#1}_{~\Lambda}${#2}} 
\newcommand{\vect}[1]{\boldsymbol{\mathbf{#1}}}
\newcommand{\etal}{\textit{et al.}}

\begin{document}

\pagestyle{plain}

\title {High-resolution hypernuclear spectroscopy at Jefferson Lab, Hall A }

\author{F. ~Garibaldi}
\affiliation{Istituto Nazionale di Fisica Nucleare, Sezione di Roma,
Piazzale A. Moro 2, I-00185 Rome, Italy}

\author{A.~Acha}
\affiliation{Florida International University, Miami, Florida 33199, USA}

\author{P.~Ambrozewicz}
\affiliation{Florida International University, Miami, Florida 33199, USA}

\author{K.A.~Aniol}
\affiliation{California State University, Los Angeles, Los Angeles
California 90032, USA}

\author{P.~Baturin}
\affiliation{Rutgers, The State University of New Jersey, Piscataway,
New Jersey 08855, USA}

\author{H.~Benaoum}
\affiliation{Syracuse University, Syracuse, New York 13244, USA}

\author{J. Benesch}
\affiliation{Thomas Jefferson National Accelerator Facility, Newport News,
Virginia 23606, USA}

\author{P.Y.~Bertin}
\affiliation{Universit\'{e} Blaise Pascal/IN2P3, F-63177 Aubi\`{e}re, France}

\author{K.I.~Blomqvist}
\affiliation{Universit\"at Mainz, Mainz, Germany}

\author{W.U.~Boeglin}
\affiliation{Florida International University, Miami, Florida 33199, USA}

\author{H.~Breuer}
\affiliation{University of Maryland, College Park, Maryland 20742, USA}

\author{P.~Brindza}
\affiliation{Thomas Jefferson National Accelerator Facility, Newport News,
Virginia 23606, USA}

\author{P.~Byd\v{z}ovsk\'y}
\affiliation{Nuclear Physics Institute, \v{R}e\v{z} near Prague, Czech
Republic}

\author{A.~Camsonne}
\affiliation{Universit\'{e} Blaise Pascal/IN2P3, F-63177 Aubi\`{e}re, France}

\author{C.C.~Chang}
\affiliation{University of Maryland, College Park, Maryland 20742, USA}

\author{J.-P.~Chen}
\affiliation{Thomas Jefferson National Accelerator Facility, Newport News,
Virginia 23606, USA}

\author{Seonho~Choi}
\affiliation{Temple University, Philadelphia, Pennsylvania 19122, USA}

\author{E.A.~Chudakov}
\affiliation{Thomas Jefferson National Accelerator Facility, Newport News,
Virginia 23606, USA}

\author{E.~Cisbani}

\author{S.~Colilli}
\affiliation{Istituto Nazionale di Fisica Nucleare, Sezione di Roma, gruppo
collegato Sanit\`a, and Istituto Superiore di Sanit\`a, I-00161 Rome, Italy}

\author{L.~Coman}
\affiliation{Florida International University, Miami, Florida 33199, USA}

\author{F.~Cusanno}
\affiliation{Istituto Nazionale di Fisica Nucleare, Sezione di Roma,
Piazzale A. Moro 2, I-00185 Rome, Italy}

\author{B.J.~Craver}
\affiliation{University of Virginia, Charlottesville, Virginia 22904, USA}

\author{G.~De~Cataldo}
\affiliation{Istituto Nazionale di Fisica Nucleare, Sezione di Bari and
University of Bari, I-70126 Bari, Italy}

\author{C.W.~de~Jager}
\affiliation{Thomas Jefferson National Accelerator Facility, Newport News,
Virginia 23606, USA}

\author{R.~De~Leo}
\affiliation{Istituto Nazionale di Fisica Nucleare, Sezione di Bari and
University of Bari, I-70126 Bari, Italy}

\author{A.P.~Deur}
\affiliation{University of Virginia, Charlottesville, Virginia 22904, USA}

\author{C.~Ferdi}
\affiliation{Universit\'{e} Blaise Pascal/IN2P3, F-63177 Aubi\`{e}re, France}

\author{R.J.~Feuerbach}
\affiliation{Thomas Jefferson National Accelerator Facility, Newport News,
Virginia 23606, USA}

\author{E.~Folts}
\affiliation{Thomas Jefferson National Accelerator Facility, Newport News,
Virginia 23606, USA}

\author{S.~Frullani}

\affiliation{Istituto Nazionale di Fisica Nucleare, Sezione di Roma,
Piazzale A. Moro 2, I-00185 Rome, Italy}


\author{O.~Gayou}
\affiliation{Massachussets Institute of Technology, Cambridge, Massachusetts
02139, USA}

\author{F.~Giuliani}
\affiliation{Istituto Nazionale di Fisica Nucleare, Sezione di Roma, gruppo
collegato Sanit\`a, and Istituto Superiore di Sanit\`a, I-00161 Rome, Italy}

\author{J.~Gomez}
\affiliation{Thomas Jefferson National Accelerator Facility, Newport News,
Virginia 23606, USA}

\author{M.~Gricia}
\affiliation{Istituto Nazionale di Fisica Nucleare, Sezione di Roma, gruppo
collegato Sanit\`a, and Istituto Superiore di Sanit\`a, I-00161 Rome, Italy}

\author{J.O.~Hansen}
\affiliation{Thomas Jefferson National Accelerator Facility, Newport News,
Virginia 23606, USA}

\author{D.~Hayes}
\affiliation{Old Dominion University, Norfolk, Virginia 23508, USA}

\author{D.W.~Higinbotham}
\affiliation{Thomas Jefferson National Accelerator Facility, Newport News,
Virginia 23606, USA}

\author{T.K.~Holmstrom}
\affiliation{College of William and Mary, Williamsburg, Virginia 23187, USA}

\author{C.E.~Hyde}
\affiliation{Old Dominion University, Norfolk, Virginia 23508, USA}
\affiliation{Universit\'{e} Blaise Pascal/IN2P3, F-63177 Aubi\`{e}re, France}

\author{H.F.~Ibrahim}
\affiliation{Cairo University, Giza 12613, Egypt}

\author{M.~Iodice}
\affiliation{Istituto Nazionale di Fisica Nucleare, Sezione di Roma Tre,
I-00146 Rome, Italy}

\author{X.~Jiang}
\affiliation{Rutgers, The State University of New Jersey, Piscataway,
New Jersey 08855, USA}

\author{L.J.~Kaufman}
\affiliation{University of Massachussets Amherst, Amherst, Massachusetts
01003, USA}

\author{K.~Kino}
\affiliation{Research Center for Nuclear Physics, Osaka
University, Ibaraki, Osaka 567-0047, Japan}

\author{B.~Kross}
\affiliation{Thomas Jefferson National Accelerator Facility, Newport News,
Virginia 23606, USA}

\author{L.~Lagamba}
\affiliation{Istituto Nazionale di Fisica Nucleare, Sezione di Bari and
University of Bari, I-70126 Bari, Italy}

\author{J.J.~LeRose}
\affiliation{Chesapeake Bay Governor's School, Tappahannock,
Virginia 222560, USA}

\author{R.A.~Lindgren}
\affiliation{University of Virginia, Charlottesville, Virginia 22904, USA}

\author{M.~Lucentini}
\affiliation{Istituto Nazionale di Fisica Nucleare, Sezione di Roma, gruppo
collegato Sanit\`a, and Istituto Superiore di Sanit\`a, I-00161 Rome, Italy}

\author{D.J.~Margaziotis}
\affiliation{California State University, Los Angeles, Los Angeles
California 90032, USA}

\author{P.~Markowitz}
\affiliation{Florida International University, Miami, Florida 33199, USA}

\author{S.~Marrone}
\affiliation{Istituto Nazionale di Fisica Nucleare, Sezione di Bari and
University of Bari, I-70126 Bari, Italy}

\author{D.G.~Meekins}
\affiliation{Thomas Jefferson National Accelerator Facility, Newport News,
Virginia 23606, USA}

\author{Z.E.~Meziani}
\affiliation{Temple University, Philadelphia, Pennsylvania 19122, USA}

\author{K.~McCormick}
\affiliation{Rutgers, The State University of New Jersey, Piscataway,
New Jersey 08855, USA}

\author{R.W.~Michaels}
\affiliation{Thomas Jefferson National Accelerator Facility, Newport News,
Virginia 23606, USA}

\author{D.J.~Millener}
\affiliation{Brookhaven National Laboratory, Upton, New York 11973, USA}

\author{T.~Miyoshi}
\affiliation{Tohoku University, Sendai, 980-8578, Japan}

\author{B.~Moffit}
\affiliation{College of William and Mary, Williamsburg, Virginia 23187, USA}

\author{P.A.~Monaghan}
\affiliation{Massachussets Institute of Technology, Cambridge, Massachusetts
02139, USA}

\author{M.~Moteabbed}
\affiliation{Florida International University, Miami, Florida 33199, USA}

\author{C.~Mu\~noz~Camacho}
\affiliation{CEA Saclay, DAPNIA/SPhN, F-91191 Gif-sur-Yvette, France}

\author{S.~Nanda}
\affiliation{Thomas Jefferson National Accelerator Facility, Newport News,
Virginia 23606, USA}

\author{E.~Nappi}
\affiliation{Istituto Nazionale di Fisica Nucleare, Sezione di Bari and
University of Bari, I-70126 Bari, Italy}

\author{V.V.~Nelyubin}
\affiliation{University of Virginia, Charlottesville, Virginia 22904, USA}

\author{B.E.~Norum}
\affiliation{University of Virginia, Charlottesville, Virginia 22904, USA}

\author{Y.~Okasyasu}
\affiliation{Tohoku University, Sendai, 980-8578, Japan}

\author{K.D.~Paschke}
\affiliation{University of Massachussets Amherst, Amherst, Massachusetts
01003, USA}

\author{C.F.~Perdrisat}
\affiliation{College of William and Mary, Williamsburg, Virginia 23187, USA}

\author{E.~Piasetzky}
\affiliation{School of Physics and Astronomy, Sackler Faculty of Exact
Science, Tel Aviv University, Tel Aviv 69978, Israel}

\author{V.A.~Punjabi}
\affiliation{Norfolk State University, Norfolk, Virginia 23504, USA}

\author{Y.~Qiang}
\affiliation{Massachussets Institute of Technology, Cambridge, Massachusetts
02139, USA}

\author{B.~Raue}
\affiliation{Florida International University, Miami, Florida 33199, USA}

\author{P.E.~Reimer}
\affiliation{Argonne National Laboratory, Argonne, Illinois 60439, USA}

\author{J.~Reinhold}
\affiliation{Florida International University, Miami, Florida 33199, USA}

\author{B.~Reitz}
\affiliation{Thomas Jefferson National Accelerator Facility, Newport News,
Virginia 23606, USA}

\author{R.E.~Roche}
\affiliation{Florida State University, Tallahassee, Florida 32306, USA}

\author{V.M.~Rodriguez}
\affiliation{University of Houston, Houston, Texas 77204, USA}

\author{A.~Saha}
\affiliation{Thomas Jefferson National Accelerator Facility, Newport News,
Virginia 23606, USA}

\author{F.~Santavenere}
\affiliation{Istituto Nazionale di Fisica Nucleare, Sezione di Roma, gruppo
collegato Sanit\`a, and Istituto Superiore di Sanit\`a, I-00161 Rome, Italy}

\author{A.J.~Sarty}
\affiliation{St. Mary's University, Halifax, Nova Scotia, Canada}

\author{J.~Segal}
\affiliation{Thomas Jefferson National Accelerator Facility, Newport News,
Virginia 23606, USA}

\author{A.~Shahinyan}
\affiliation{Yerevan Physics Institute, Yerevan, Armenia}

\author{J.~Singh}
\affiliation{University of Virginia, Charlottesville, Virginia 22904, USA}

\author{S.~\v{S}irca}
\affiliation{Faculty of Mathematics and Physics, University of Ljubljana, 
Slovenia}

\author{R.~Snyder}
\affiliation{University of Virginia, Charlottesville, Virginia 22904, USA}

\author{P.H.~Solvignon}
\affiliation{Temple University, Philadelphia, Pennsylvania 19122, USA}

\author{M.~Sotona}
\affiliation{Nuclear Physics Institute, \v{R}e\v{z} near Prague, Czech
Republic}

\author{R.~Subedi}
\affiliation{Kent State University, Kent, Ohio 44242, USA}

\author{V.A.~Sulkosky}
\affiliation{College of William and Mary, Williamsburg, Virginia 23187, USA}

\author{T.~Suzuki}
\affiliation{Tohoku University, Sendai, 980-8578, Japan}

\author{H.~Ueno}
\affiliation{Yamagata University, Yamagata 990-8560, Japan}

\author{P.E.~Ulmer}
\affiliation{Old Dominion University, Norfolk, Virginia 23508, USA}

\author{G.M.~Urciuoli}
\affiliation{Istituto Nazionale di Fisica Nucleare, Sezione di Roma, 
P.zza Aldo Moro 2, Rome, Italy}

\author{E.~Voutier}
\affiliation{LPSC, Universit\'e Joseph Fourier, CNRS/IN2P3, INPG, F-38026
Grenoble, France}

\author{B.B.~Wojtsekhowski}
\affiliation{Thomas Jefferson National Accelerator Facility, Newport News, 
Virginia 23606, USA}

\author{X.~Zheng}
\affiliation{Argonne National Laboratory, Argonne, Illinois 60439, USA}

\author{C.~Zorn}
\affiliation{Thomas Jefferson National Accelerator Facility, Newport News, 
Virginia 23606, USA}

\collaboration{Jefferson Lab Hall A Collaboration}
\noaffiliation
\date{\today}

\begin{abstract}

The experiment E94-107  in Hall A at Jefferson Lab started  a systematic study
of high resolution hypernuclear spectroscopy in the 0p-shell region of nuclei 
such as the hypernuclei produced in electroproduction on $^{9}$Be, $^{12}$C and 
$^{16}$O targets. In order to increase counting rates and provide unambiguous 
kaon identification two superconducting septum magnets and a ring-imaging 
\v{C}herenkov detector  were added to the Hall A standard equipment.  The 
high-quality beam, the good spectrometers and the new experimental devices 
allowed us to obtain very good results. For the first time, measurable strength 
with sub-MeV energy resolution was observed for the core-excited states of 
\lam{12}{B}. A high-quality \lam{16}{N} hypernuclear spectrum was likewise
obtained. A first measurement of the $\Lambda$ binding energy for \lam{16}{N}, 
calibrated against the elementary reaction on hydrogen, was obtained with 
high precision, $13.76\pm 0.16$ MeV. Similarly, the first \lamb{9}{Li} 
hypernuclear spectrum shows general agreement with theory (distorted-wave
impulse approximation with the SLA and BS3 electroproduction models and 
shell-model wave functions). Some disagreement exists with respect to the 
relative strength of the states making up the first multiplet. A $\Lambda$ 
separation energy of 8.36 MeV was obtained, in agreement with previous 
results. It has been shown that the electroproduction  of hypernuclei can 
provide information complementary to that obtained with hadronic probes 
and the $\gamma$-ray spectroscopy technique.

\end{abstract}

\maketitle

\section{Introduction}

The physics of hypernuclei, multibaryonic systems with non-zero strangeness, 
is an important branch of contemporary nuclear physics at low energy 
(structure, energy spectra and weak decays of hypernuclei) as well as at 
intermediate energy (production mechanism)~\cite{gal16}. The $\Lambda$  
hypernucleus is  a long-lived baryonic system ($ t=10^{-10} s$) and provides 
us with a variety of nuclear phenomena. The hyperon inside an ordinary 
nucleus is not affected by the Pauli principle and can penetrate deeply 
inside the nucleus permitting measurements of the system response to the 
stress imposed on it. The study of its propagation can reveal configurations 
or states not seen in other ways. The study also gives important insight 
into the structure of ordinary nuclear matter.

An understanding of baryon-baryon interactions is fundamental in order 
to understand our world and its evolution. However, our current knowledge is 
limited at the level of strangeness zero particles (p and n). Hence, studying 
the hyperon-nucleon (YN) and hyperon-hyperon (YY) interactions is very 
important in order to extend our knowledge and seek a unified description of 
them. Since very limited information can be obtained from elementary 
hyperon-nucleon scattering, hypernuclei are unique laboratories for studying
the $\Lambda N$ interaction~\cite{PTP2010}. In fact, an effective $\Lambda N$ 
interaction can be determined from hypernuclear spectra obtained from various 
reactions and can be used to discriminate between different YN  potentials 
employed to carry out ab initio many-body calculations~\cite{lonardoni13}.

Until now a large body of data came from two types of highly complementary 
hypernuclear spectroscopy techniques: reaction based spectroscopy with 
hadron probes and $\gamma$-ray spectroscopy~\cite{hashtam06}. Reaction 
spectroscopy, which directly populates hypernuclear states, reveals the 
level structure in the $\Lambda$ bound region and can even study excited 
states between the nucleon emission threshold and the $\Lambda$ emission 
threshold. It provides information on $\Lambda$ hypernuclear structure 
and the $\Lambda$ emission threshold. The information on $\Lambda$ 
hypernuclear structure and the $\Lambda N$ interaction is obtained through the 
determination of the hypernuclear masses, spectra, reaction cross sections, 
angular distributions etc. Moreover, precise measurements of the production 
cross 
sections provide information on the hypernuclear production mechanism and 
the dynamics of the elementary-production reaction. $\gamma$-ray 
spectroscopy achieves ultra-high resolution (typically a few keV). 
It is a powerful tool for investigation of the spin dependent part of 
the $\Lambda N$ interaction, which requires precise information on the 
level structure of hypernuclei. Both these powerful techniques have 
limitations, first limited energy resolution and small spin-flip amplitudes, 
and second the access only to hypernuclear states below 
nucleon emission threshold. 

Experimental knowledge can be greatly improved using electroproduction of 
strangeness characterized by large three-momentum transfer 
($\sim250$ MeV/c), large angular momentum transfer $\Delta J$, and strong 
spin-flip terms, even at zero production angles~\cite{hashtam06}. 
Moreover, the $K^+\Lambda$ pair production occurs on a proton in contrast 
to a neutron in ($K^-,\pi^-$) or ($\pi^+,K^+$) reactions making possible the 
study of different hypernuclei and charge-dependent effects from a comparison 
of mirror hypernuclei (charge-symmetry breaking). The hypernuclear 
$\gamma$-ray measurements give extremely high-precision 
energy-level spacings, while the precision of the energy levels given by 
the ($e, e'K^+$) reaction spectroscopy can potentially be a few hundreds 
of keV, which is more than an order of magnitude worse than $\gamma$-ray 
measurement. However, the advantage of being able to simultaneously observe
more complete structures, as well as to provide precise absolute binding 
energy is obvious. For transitions with energy larger than 1 MeV, a Ge 
detector efficiency decreases quickly, and thus statistics becomes a major 
problem for the current $\gamma$-ray spectroscopy program using the Ge 
detector technique. 

Even though plans for various new hypernuclear physics studies at other 
facilities exist, the precision, accurate mass spectroscopy from the JLab 
program has a unique position, in addition to the clearly known 
common advantages of electro-production (such as the size of momentum 
transfer allowing large angular momentum transfer, extra spin transfer 
from the virtual photon, converting a proton into a
$\Lambda$ to study neutron-rich hypernuclei). 

The E94-107 experiment in Hall A at Jefferson Lab ~\cite{proposal94} started 
a systematic study of high-resolution hypernuclear spectroscopy on p-shell 
targets, specifically $^9$Be~\cite{urciuoli15}, $^{12}$C~\cite{iodice07} and 
$^{16}$O~\cite{cusanno09}. Moreover a study of the elementary reaction on 
a proton was performed.

This paper describes the experimental apparatus, the theoretical models, 
the results obtained and the physics information extracted from them.

\section{The experiment}
\label{sec:experiment}

Hall A at Jefferson Lab is well suited to perform $(e,e'K^+)$ experiments. 
Scattered electrons can be detected in the high-resolution spectrometer 
(HRS) electron arm while coincident kaons are detected in the HRS hadron 
arm~\cite{alcorn04}. The disadvantage of smaller electromagnetic cross 
sections with respect to hadron-induced reactions is partially 
compensated for by the high current, high duty cycle, and high energy 
resolution capabilities of the beam. The detector packages for the electron 
and hadron spectrometers are almost identical, except for the particle 
identification (PID) systems discussed later~\cite{alcorn04}.

\begin{figure} 
\centering
\includegraphics[width=7.5cm]{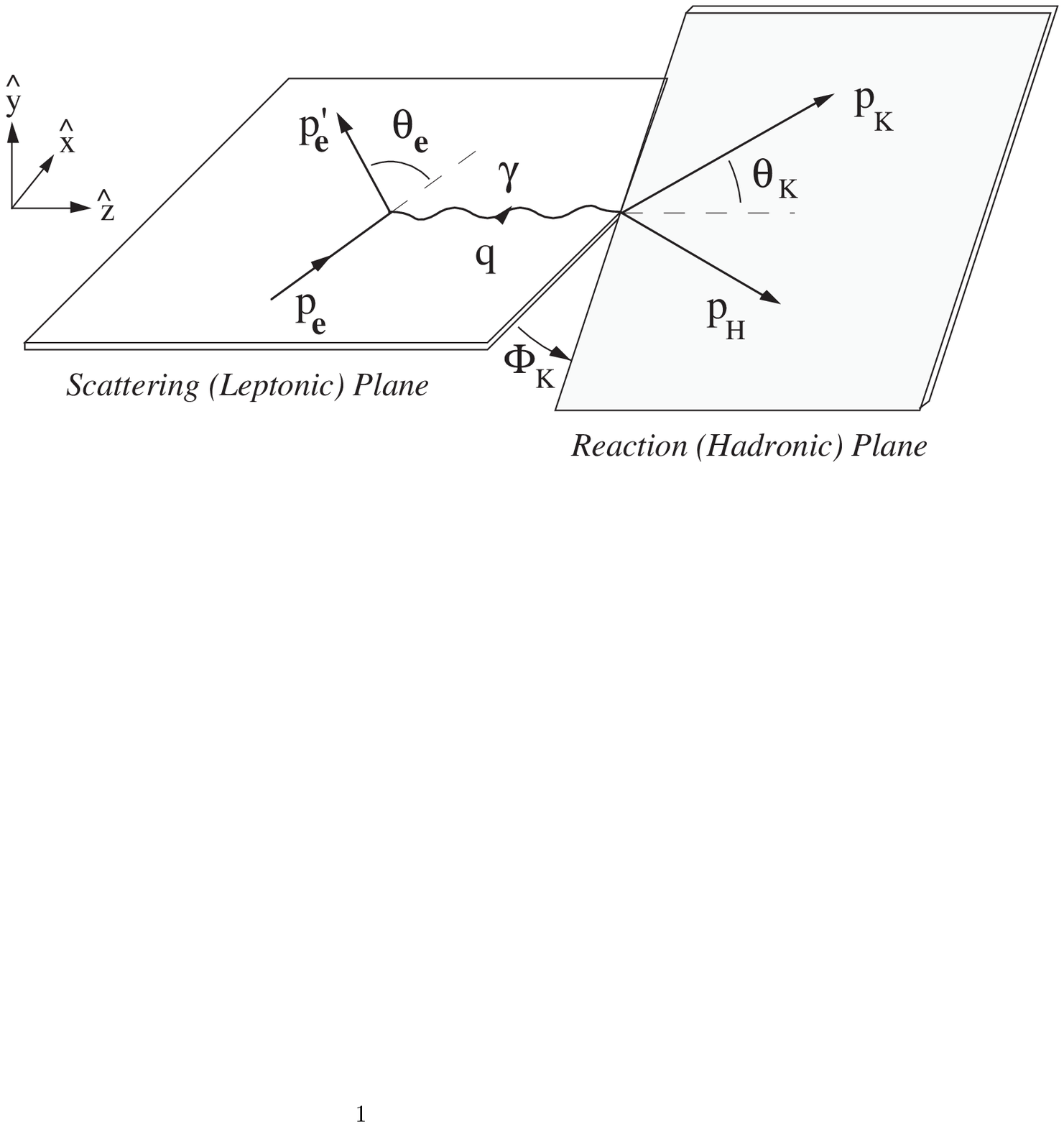}
\caption{Kinematics of hypernuclear electroproduction in the laboratory frame.}
\label{fig:kinematics}
\end{figure}

The kinematics for the three experiments is shown in Fig.~\ref{fig:kinematics} 
and values are given in Table~\ref{tab:kinematics}. The beam and 
final electron energies are denoted $E_i$ and $E_f$, respectively, and the 
electron ($\theta_e$), kaon ($\theta_{Ke}$), and photon ($\theta_{\gamma e}$) 
angles are measured with respect to the beam direction. The virtual photon 
energy, transverse polarisation, and flux factor are denoted as 
$E_\gamma$, $\epsilon$, and $\Gamma$. The kaon momentum $p_K$ changes a 
little bit due to a different hypernucleus mass for the excited states. 
A coplanar experimental setup was chosen with the kaon azimuthal angle 
$\Phi_K=180^\circ$. Then the kaon lab angle with respect to the photon 
direction is $\theta_K=\theta_{Ke}-\theta_{\gamma e}$, see 
Fig.~\ref{fig:kinematics}.     
%
%
%
%
\begin{table*}[bt]
\caption{Kinematics in the laboratory frame for the three experiments.
\label{tab:kinematics}}
\begin{ruledtabular}
\begin{tabular}{ccccccccccc} 
Target & $E_i$  & $E_f$ & $\theta_e$ & $\theta_{Ke}$ & $E_\gamma$ & 
$\theta_{\gamma e}$ & $Q^{2}$ & $\epsilon$ & $\Gamma$ &   $p_K$   \\
       & [GeV]  & [GeV] &   [deg]    &   [deg]       &  [GeV]     &
   [deg]            & [GeV$^2$] &      &  [(GeV sr)$^{-1}$]& [GeV]\\     
\hline
$^{9}Be$ & 3.77 & 1.56  &   6        &    6        &   2.21 & 
    4.20            & 0.0644  &  0.703     & 0.0174   &   1.96     \\
$^{12}C$ & 3.77 & 1.56  &   6        &    6        &   2.21 &
    4.20            & 0.0644  &  0.703     & 0.0174   & 1.95--1.96 \\
$^{16}O$ & 3.66 & 1.45  &   6        &    6        &   2.21 & 
    3.91            & 0.0581  &  0.682     & 0.0172   & 1.95--1.97 \\
\end{tabular}
\end{ruledtabular}
\end{table*}

The reasons for this choice were the following. The momentum transfer 
to the hypernucleus in the electroproduction is rather large (350 MeV/c) 
and decreases steadily with increasing energy of the virtual photon 
($E_\gamma=E_e-E_{e'}$) while the elementary electroproduction cross section, 
with the kaon detected at forward angles, is almost constant for 
$E_\gamma$=1.2-2.2 GeV. The momentum transfer for forward kaon scattering 
angles falls from 330 MeV/c at $E_\gamma$ = 1.2 GeV to 250 MeV/c at 
$E_\gamma$ = 2.5 GeV, so that higher energies are preferable. 
Moreover, because the cross section depends strongly on Q$^{2}$  
(through the virtual photon flux as determined by the electron
kinematics), the measurements have to be made at low Q$^{2}$ to get reasonable 
counting rates . Hence, the electron scattering angle 
must be small, and the kaon angle close to the virtual photon direction
in order to minimize the momentum transfer. Moreover, due to the long flight 
path in the HRS spectrometer, to keep a reasonable 
kaon survival fraction the kaon momenta must be fairly high.

Good energy resolution together with a low level of background is 
mandatory for this experiment. The energy resolution depends on the momentum 
resolution  of the HRS spectrometers, on the straggling
and energy loss in the target, and on the beam energy spread. 
A momentum resolution of the system (HRS's + septum magnets)
of $\Delta p / p = 10^{-4}$ (FWHM) and a beam energy spread as 
small as $6\times 10^{-5}$ (FWHM) are necessary to be able to get 
an excitation energy resolution of 700 keV or less. A very good PID
system is needed to guarantee a low level of background. 

\subsection{The beam} 

\subsubsection{Beam monitors}
\label{subsec:beam-mon}

E94-107 desired a continuous wave, 3.66~GeV, \mbox{100 $\mu$A} electron beam 
with very small energy spread and vertical spot size (energy spread 
$\sigma\leq 3 \times 10^{-5}$, spot size $\sigma\leq~100~\mu$m).
With some effort, the CEBAF staff were able to achieve these requirements. 
The absolute value of the beam energy was measured using the Arc method
(see Sec.~\ref{subsec:arc}). The beamline is segmented into several sections 
isolated by vacuum valves. The beam diagnostic elements consist of 
transmission-line position monitors, current monitors, superharps, viewers, 
loss monitors, and optical transition radiation (OTR) viewers. 
Drifts in the central beam energy were monitored using the so-called 
``Hall A Tiefenback Energy'' value (see Sec.~\ref{subsec:tiefenback}). 
The beam spot size and the energy spread were continuously monitored
using a Synchrotron Light Interferometer (SLI)~\cite{chao04} (see
Sec.~\ref{subsec:sli}).

\subsubsection{The Arc method}
\label{subsec:arc}

The Arc method determines the energy by measuring the deflection of the 
beam in the arc section of the beamline. The nominal bend angle of the beam
in the arc section is 34.3$^\circ$. The measurement is made when the beam 
is tuned in dispersive mode in the arc section. The momentum of the beam 
is then related to the field integral of eight dipoles and the net bend 
angle through the arc section \cite{alcorn04}. 
The method consists of two simultaneous measurements, one for the magnetic
field integral of the bending elements, and one for the actual bend angle of 
the arc.

\subsubsection{Hall A Tiefenback}
\label{subsec:tiefenback}

``Hall A Tiefenback'' is a beam diagnostic tool developed by Mike Tiefenback 
of the JLab Accelerator Scientific staff that uses position monitors and
magnet settings in the arc leading to Hall A to monitor relative shifts
in the beam energy. 

\subsubsection{Synchrotron light interferometer (SLI)}
\label{subsec:sli}

An SLI has been used at Jefferson Lab in
order to measure small beam sizes below the diffraction limit.
The device is not invasive and can monitor the profile of electron beam. 
The SLI at Jefferson Lab is a wave-front division interferometer that uses 
polarized quasi-monochromatic synchrotron light. 
The syncrotron light generated by the electron beam in a dipole magnet is
extracted through a quartz window. After this window, the light is optically  
shielded until it reaches a CCD video camera connected to the image processor. 
An optical system, comprising two adjustable 45$^\circ$ mirrors and a 
diffraction limited doublet lens, produces an interferogram.  
The basic parameter to calculate the beam size is the visibility, V, of the
interference pattern. The visibility is estimated from the intensities of the
first (central) maximum ($I_{max}$) and minima ($I_{min}$) of the 
interferogram 
\begin{equation}
V = \frac{I_{max} - I_{min}}{I_{max} + I_{min}} \ .
\end{equation}

Assuming a Gaussian beam shape, the spread of the beam can be calculated.

\subsection{Spectrometers and septum magnets}
\label{subsec:spect}

The standard equipment HRS pair~\cite{alcorn04} in Hall A was designed to 
deliver the required momentum resolution. However, because the hypernuclear
cross section falls rapidly with increasing angle (momentum transfer),
the minimum angles with respect to the beamline 
of $12.5^\circ$ were too large. This shortcoming was 
mitigated by the introduction of a pair of superconducting septum magnets 
providing a $6.5^\circ$ horizontal bend each. By moving the target 
postion 80~cm upstream and inserting the septum magnets on either side of 
the beamline, the HRS pair at $12.5^\circ$ on either side of the beamline
is able to detect kaons and electrons at $6^\circ$.
This new spectrometer configuration (Septum + HRS) provides a general
purpose device that extends the HRS features to small scattering angles  
while preserving the spectrometer optical performance~\cite{urciuoli01}.

The septum magnets have to fulfill the following requirements. They must 
match the entrance optics of the HRS spectrometers for a pivot displaced 
by $0.8$ m downstream of the target in an angular range of 6$^\circ$ to 
12.5$^\circ$. The septa must bend 4 GeV/c particles of any polarity at 
any angle from 6$^\circ$ to 12.5$^\circ$ and match the HRS optics from 
12$^\circ$ to 24$^\circ$. The unique location of the septa at the match point, 
the short space between the displaced scattering chamber and the first HRS 
quadrupole, and the proximity to the outgoing electron beam, imposes severe 
space constraints. These lead to a requirement for a relatively high central 
field of 4.2 T. A 2.77 Tm field integral and an aperture centered on 6$^\circ$
with an acceptance of $\pm 24 \textrm{mrad (horizontal)}\times \pm 54
\textrm{mrad (vertical)}$ is needed. The requirement for less than 0.08 Tm 
field 
integral along the exit electron beam leads directly to a super-conducting 
window frame coil in a C type iron geometry with a relatively high, 
25 kA/cm$^{2}$, current density. The TOSCA field maps were used as input
to a ray-tracing code and simulations of the spectrometer performance
were made to prove that the ``as designed'' magnetic fields satisfied
the optics requirements. The septum field quality is determined by an 
experimental resolution requirement of overall $\Delta p / p = 10^{-4}$ and 
the use of the optics simulations to verify that the overall magnetic system 
is consistent with preservation of spectrometer performance. For details of 
the design and construction of the septum magnets see Ref.~\cite{urciuoli01}.

A very nice feature of the septum magnet setup was that the two arms were 
essentially independent and could be tuned and optimized separately.
Due to their small bend angle and relatively short length (80 cm) the 
septum magnets made only a modest perturbation on the standard HRS optics 
that was easily corrected by a small tuning of the three quadrupoles in 
each arm. 

\subsection{Targets}
\label{subsec:targets}

A standard cryogenic target~\cite{alcorn04} was used for the study of the
elementary reaction. Standard solid targets (100 mg/cm$^2$) were used for 
$^{9}$Be and $^{12}$C.  A waterfall target system was used for experiments on 
$^{16}$O~\cite{garibaldi92}. This target has also been used for studying the 
elementary reaction.

\subsection{Waterfall target}
\label{subsec:waterfall}

\begin{figure} 
\centering
\includegraphics[width=9cm, clip]{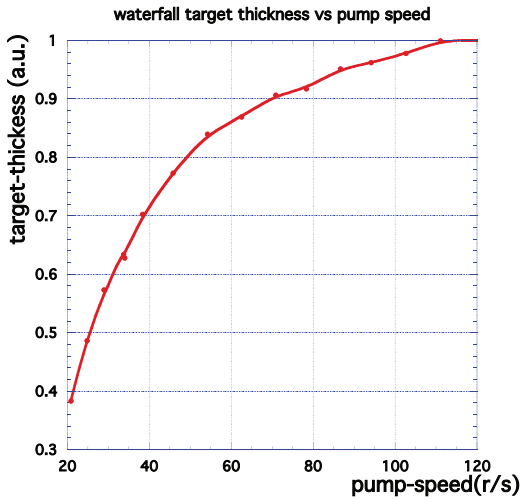}
\caption{Target thickness vs pump speed}
\label{fig:water1}
\end{figure}

\begin{figure} 
\centering
\includegraphics[width=6.0cm,angle=270]{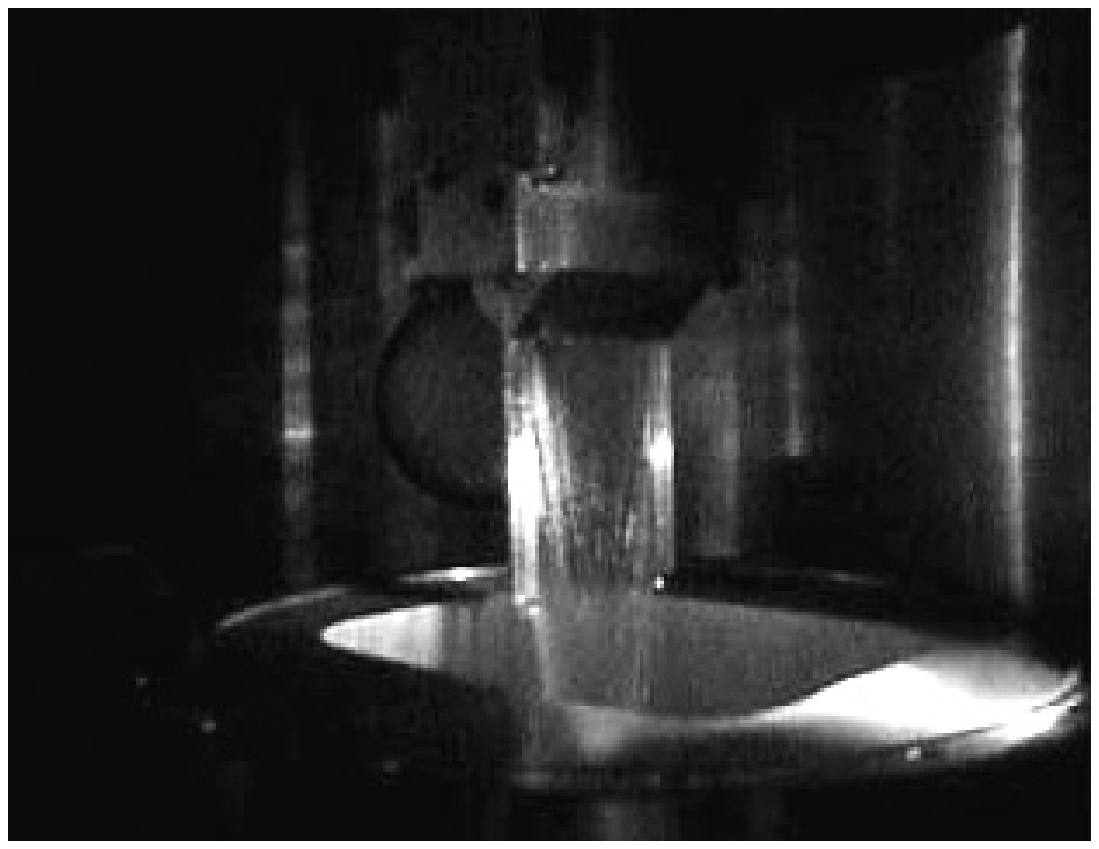}
\caption{View of the target cell with the waterfall}
\label{water}
\end{figure}

The waterfall target system provides a target for experiments on $^{16}$O. 
Using a waterfall for oxygen experiments has many advantages. Pure oxygen 
is difficult to handle, as it is highly reactive. The use of other oxygen 
compounds requires additional measurements to subtract the non-oxygen 
background, whereas the hydrogen in water can be used for calibration purposes.
The technique of using continuously flowing water as an electron-scattering 
target was first developed by Voegler and Friedrich~\cite{voegler82}, 
and later refined by Garibaldi et al.~\cite{garibaldi92}. The waterfall foil 
is produced in a cell mounted in the standard scattering chamber. Water 
forced through slits forms a flat rectangular film which is stable as a 
result of surface tension and adherence to stainless steel poles (see 
Fig.~\ref{water}). The water, continuously pumped from a reservoir, goes 
through a heat exchanger into the target zone and then back into the 
reservoir. All parts in contact with the  water are made of stainless 
steel. Once the target is formed the thickness increases with the pump 
speed up to a maximum value that depends on the dimension of the slits 
and the stainless steel poles~\cite{garibaldi92}. A factor of  $\sim3$ 
magnification is possible (see Fig.~\ref{fig:water1}).

%
%

The target thickness stability is monitored by continuously measuring the 
pump speed, the flow rate and the electron rate. The target is designed to 
stay at a fixed angular position. Care has to be taken in choosing the 
window material because of the risk of melting for high beam currents
(50 $\mu$A in this case). The entrance and exit windows are circular 
(30 mm in diameter) and made of Be (75 $\mu$m thick). Because Be is highly 
toxic, it has been plated with 13 $\mu$m of Ni and a monolayer of Au (which 
also serves to improve heat conductivity). Under the cell a target frame
 holds up to five solid targets. A target position can be selected remotely 
by a mechanical system driven by stepping motors and controlled by absolute 
encoders whose precision is 0.1 mm.

The presence of the hydrogen has many advantages. In particular, it permits 
a calibration of the missing-mass scale and thus an accurate measurement of 
the $\Lambda$-binding energy in the hypernucleus. The  $\Lambda$-peak 
position from the reaction on hydrogen can be obtained using the nominal 
central values for the kinematic variables, and then constrained to be zero 
by applying a small shift to the energy of the beam (the quantity with the 
largest uncertainty). This shift is common to reactions on hydrogen and 
oxygen and therefore its uncertainty does not affect the determination of 
the binding energies of the \lam{16}{N} levels.

%
%

\subsection{Detector package}
\label{subsec:detector}

The detector packages of the two spectrometers are designed to perform 
various functions that include triggering to activate the data-acquisition 
electronics, collecting tracking information (position and direction), 
precise timing for time of flight measurements and coincidence determination, 
and identification of the scattered particles. The timing information as well 
as the main trigger is provided from scintillators. The particle 
identification is obtained from threshold \v{C}herenkov type detectors (aerogel 
and gas) and lead-glass shower counters. The main part of the detector 
package in the two spectrometers (trigger scintillators and vertical drift
chambers) is identical. For details, see~\cite{alcorn04}.

\subsubsection{Tracking}
\label{subsec:tracking}

The HRS's have small acceptance. Tracking information is provided by a pair 
of vertical drift chambers in each spectrometer. A simple analysis 
algorithm is sufficient because multiple tracks are rare. 

\subsubsection{Triggering}
\label{subsec:triggering}

There are two primary trigger scintillator planes (S1 and S2), separated by 
a distance of about 2 m. The time resolution per plane is approximately 
0.30 ns. For experiments which need a high hadron trigger efficiency, an 
additional scintillator trigger counter (S0) can be installed. The 
information from the gas \v{C}herenkov counter can be added into the trigger. 
A coincidence trigger is made from the time overlap of the two spectrometer 
triggers in a logical AND unit. The various trigger signals go to the 
trigger supervisor module which starts the data-aquisition readout.

\subsubsection{Particle IDentification (PID)}
\label{subsec:pid}

\paragraph{Time Of Flight (TOF)}
\label{subsec:tof}

The long path from the target to the HRS focal plane (25 m) allows accurate 
time-of-flight identification in coincidence experiments if the accidental 
rate is low. After correcting for differences in trajectory lengths, a TOF 
resolution of $\sim 0.5$ ns ($\sigma$) is obtained. The time-of-flight between 
the S1 and S2 planes is also used to measure the speed of particles, 
$\beta$, with a resolution of 7\% ($\sigma$).

\paragraph{Shower Counters}
\label{subsec:shower}

Two layers of shower detectors~\cite{alcorn04} are installed in each HRS. 
The blocks in both layers in HRS-L and in the first layer in HRS-R are 
oriented perpendicular to the particle tracks. In the second layer of HRS-R, 
the blocks are parallel to the tracks. Typical pion rejection ratios of 
500:1 are achieved using two-dimensional cuts of the energy deposited in 
the front layer versus the total energy deposited.

\paragraph{Gas \v{C}herenkov}
\label{subsec:cherenkov}

A gas \v{C}herenkov detector filled with CO$_2$ at atmospheric 
pressure~\cite{iodice98} is mounted between the trigger scintillator planes 
S1 and S2. The detector allows an electron identification with 99\% efficiency 
and has a threshold for pions at 4.8 GeV/c. The detector has ten light-weight 
spherical mirrors~\cite{cisbani03}  with 80 cm focal length, each viewed by a 
photo-multiplier tube (PMT) (Burle 8854). The focusing of the \v{C}herenkov 
ring onto a small area of the PMT photocathode leads to a high current 
density near the anode. To prevent a non-linear PMT response even in the 
case of few photoelectrons requires a progressive high-voltage divider. 
The length of the particle path in the 
gas radiator is 130 cm for the gas \v{C}herenkov in the HRS-R, leading to an 
average of about twelve photoelectrons. In the HRS-L, the gas \v{C}herenkov 
detector in its standard configuration has a pathlength of 80 cm, yielding 
seven photoelectrons on average. The total amount of material in the particle 
path is about 1.4\% X0. Because of its reduced thickness, the resolution 
in HRS-L is not as good as that of the shower detector in HRS-R. The  
combination of the gas \v{C}herenkov and shower detectors provides a pion 
suppression above 2 GeV/c of a factor of $2\times 10^{-5}$, with a 98\% 
efficiency for electron selection in the HRS-R.

\paragraph{Aerogel \v{C}herenkov}
\label{subsec:aerogel}

There are two aerogel \v{C}herenkov counters available with different indices 
of refraction, which can be installed in either spectrometer and allow a 
clean separation of pions, kaons, and protons over the full momentum range 
of the HRS spectrometers. The aerogel is continuously flushed with dry CO$_2$ 
gas. The two counters (A1 and A2) are diffusion-type aerogel counters. A1 
has 24 PMT's (Burle 8854). The 6 cm thick aerogel radiator used in A1 has 
a refractive index of 1.015, giving a threshold of 2.84 (0.803) GeV/c for 
kaons (pions). The average number of photoelectrons for GeV electrons in 
A1 is $\sim 8$.  The 9 cm thick aerogel radiator used in A2 has a refractive 
index of 1.055, giving a threshold of 1.55 (2.94) GeV/c for kaons (protons).  
It is viewed by 26 PMT's (Burle 8854). Trigger logic is used to require that 
A1 not fire (e.g., rejecting pions) but that A2 does fire (requiring kaons).  
Rejection factors of 70:1 for rejecting pions and $>$ 60:1 for protons 
were achieved using the aerogels in the hardware trigger.

\paragraph{Ring Imaging \v{C}herenkov Detector (RICH)}
\label{subsec:rich}

\begin{figure} 
\centering
\includegraphics[width=6cm,angle=90]{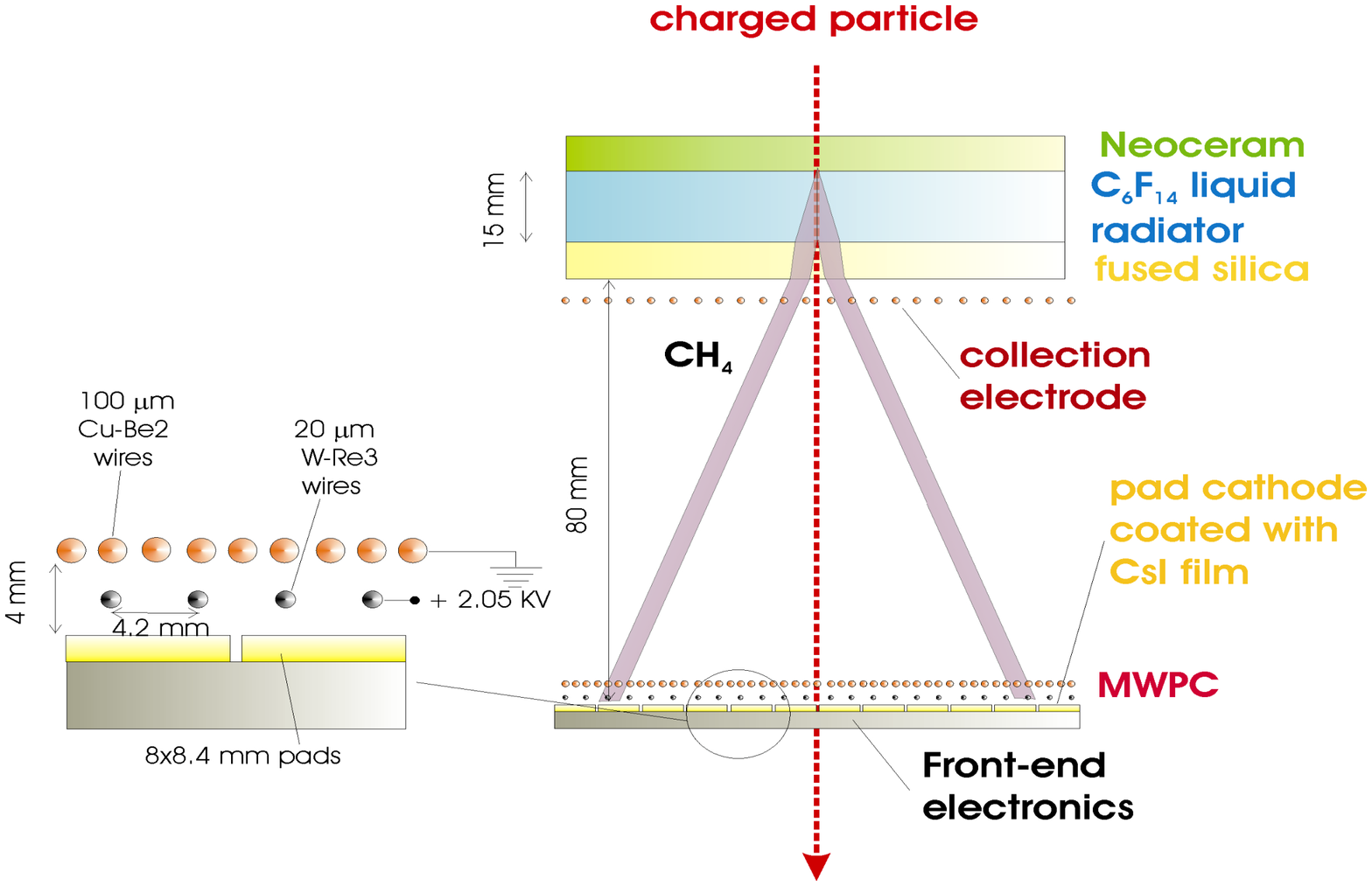}
\caption{Layout and working principle of the freon CsI
proximity focusing RICH.}
\label{rich01}
\end{figure}

In order to reduce the background level in produced spectra, a very efficient 
PID system is necessary for unambiguous kaon identification. In the electron 
arm, the gas \v{C}herenkov counters give pion rejection ratios up to $10^{3}$. 
The dominant background (knock-on electrons) is reduced by a further 2 orders 
of magnitude by the lead glass shower counters, giving a total pion rejection 
ratio of $10^{5}$. The standard PID system in the hadron arm is composed of 
two aerogel threshold \v{C}herenkov counters~\cite{lagamba01,alcorn04} 
(n$_1 = 1.015$,  n$_2 = 1.055$).  Charged pions (protons) with momenta around 
2 GeV/c are above (below) the \v{C}herenkov light emission threshold. Kaons 
emit \v{C}herenkov light only in the detector with the higher index of 
refraction. Hence, a combination of the signals from the two counters should 
distinguish among the three species of hadrons. However, due to possible 
inefficiencies and delta-ray production, the identification of kaons could have
significant contamination from pions and protons resulting in an unacceptable 
signal-to-noise ratio in the physics spectra. For these reasons the need for 
an unambiguous identification of kaons has driven the design, construction, 
and installation of a RICH detector in the hadron 
HRS focal plane detector package. The layout of the RICH is conceptually 
identical to the ALICE HMPID design~\cite{alice}. A detailed description of 
the layout and the performance of the RICH detector can be found 
in~\cite{iodice05,garibaldi03,cusanno03}. It uses a proximity-focusing geometry 
(no mirrors involved), a CsI gaseous photocathode, and a 15 mm thick liquid 
perfluorohexane radiator~\cite{alice}. Fig. ~\ref{rich01} shows the layout 
%
%
and the working principle of the adopted solution. The \v{C}herenkov photons, 
emitted along a conic surface in the radiator, are refracted by the
perfluoro-hexane (C6F14)-quartz-methane interface and strike a cathode plane 
segmented in small pads after traversing a proximity gap of 10 cm
filled with pure methane. The photon detector is made of a multi-wire 
proportional chamber (MWPC), with one cathode formed by the pad planes 
allowing for the 2-dimensional localization of the photon hit. Three 
photocathode modules of dimensions $640\times 400$ mm$^2$ segmented in 
$8\times 8.4$ mm$^2$ pads are assembled together for a total length of 
1940 mm. The pad planes are covered by a thin (300 nm) substrate of CsI 
which acts as photon converter. The emitted photoelectron is accelerated by 
an electrostatic field between the pad plane (the cathode of the MWPC) and an
anode wire plane at a distance of 2mm from it. The induced charge on the 
pads is read out by a front-end electronics based on GASSIPLEX chips.
A total number of 11520 pad channels are read out by CAEN VME V550 Flash 
ADC modules~\cite{alice}.

\paragraph{Performance}

\begin{figure} 
\centering
\includegraphics[width=6cm,angle=90]{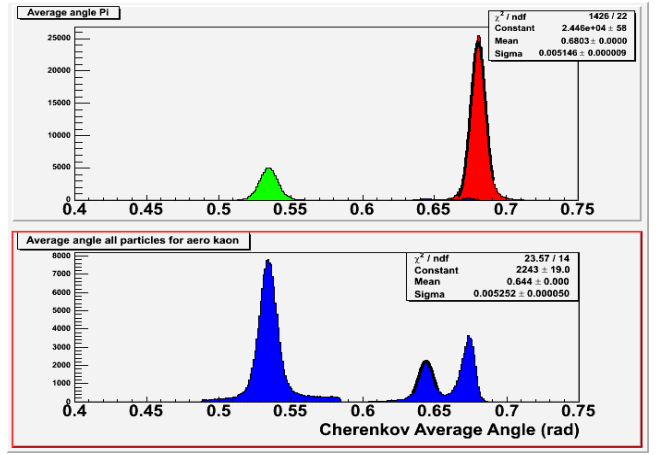}
\caption{\v{C}herenkov angle distributions for protons
(thetaC = 0.54 rad), kaons (thetaC = 0.64 rad) and pions
(thetaC = 0.68 rad). The aerogel particle selection has been used
as explained in the text}.
\label{rich02}
\end{figure}

The RICH worked successfully during the experiment~\cite{cusanno} 
where hadrons were detected in the momentum range $p\!=\!1.96\pm 0.1$ GeV/c. 
The average number of photoelectrons detected for pions is $N_\pi\!=\!13$
while for protons $N_p\!=\!8$, their ratio being in perfect agreement with 
the expected ratio of produced photons at 1.96 GeV/c. In Fig.~\ref{rich02} 
the reconstructed \v{C}herenkov angle distributions are reported. In the top 
panel the angular distributions have been obtained using samples of $\pi^+$, 
$K^+$ and $p$ as selected by the two aerogel counters. The kaon selected 
sample is practically not visible due to the very high pion to kaon ratio. 
For the dominant contribution of pions the obtained angle resolution
is $\sigma_c\!= \!5$ mrad, in agreement with Monte Carlo 
simulations~\cite{cusanno}. The kaon contribution is shown in the 
bottom panel where a large sample of aerogel kaon selected events has been 
used. The reconstructed \v{C}herenkov angle variable can be clearly used to get 
rid of the pion and proton contamination. With a resolution $\sigma_c\!=\!5$
mrad the separation between pions and K is about $6\sigma$. The performance 
reported here has been obtained with a measured quantum efficiency of
about 25\% at 160 nm~\cite{cusanno03}. The RICH pion rejection factor can be 
estimate to be $\sim 1000$ from the pion peak content reduction factor.

%
%

\paragraph{The Evaporator}
\label{subsec:evaporator}

\begin{figure} 
\centering
\includegraphics[width=8cm]{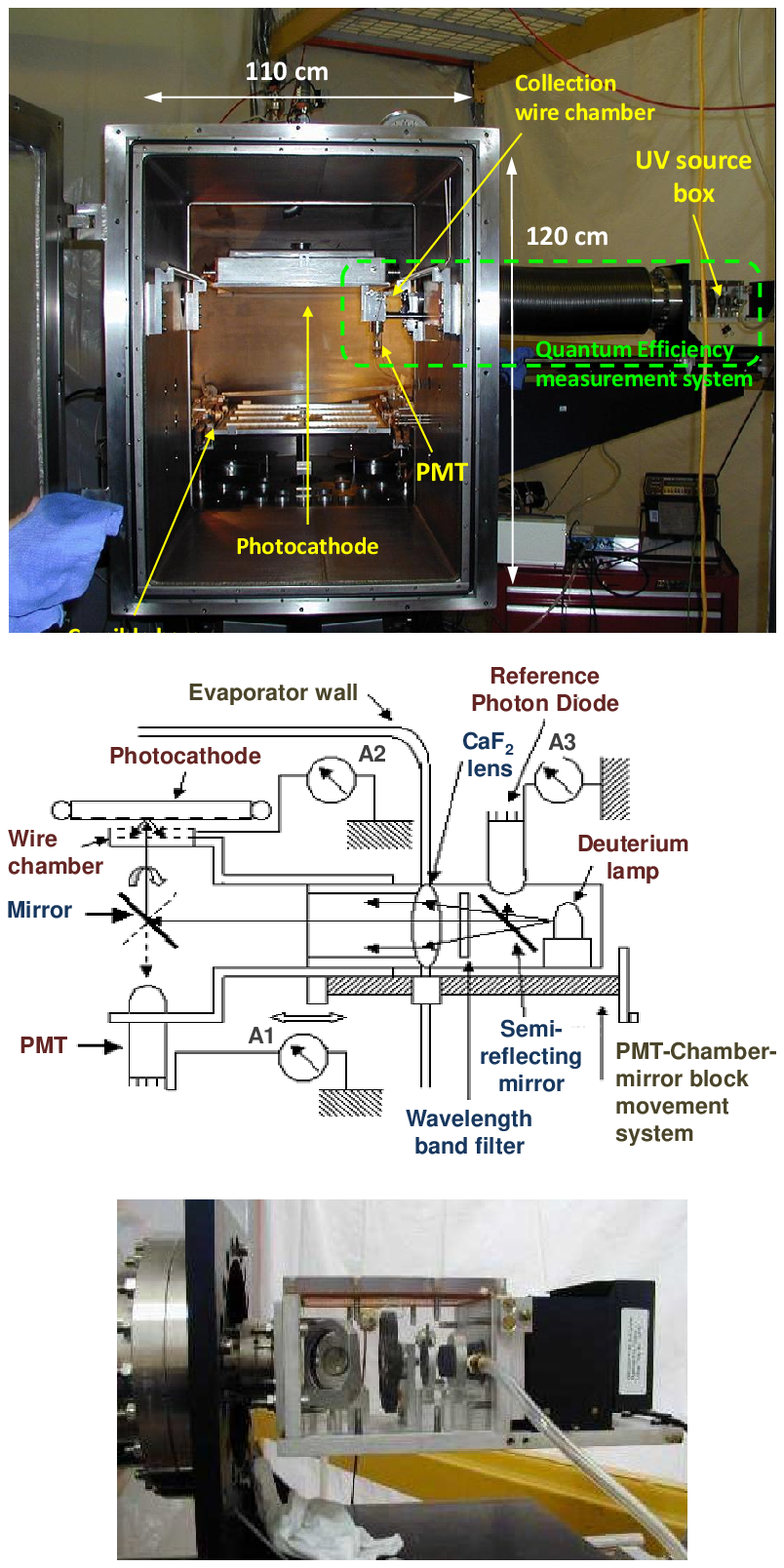}
\caption{The CsI evaporator system}
\label{evap1}
\end{figure}

\begin{figure} 
\centering
\includegraphics[width=8cm]{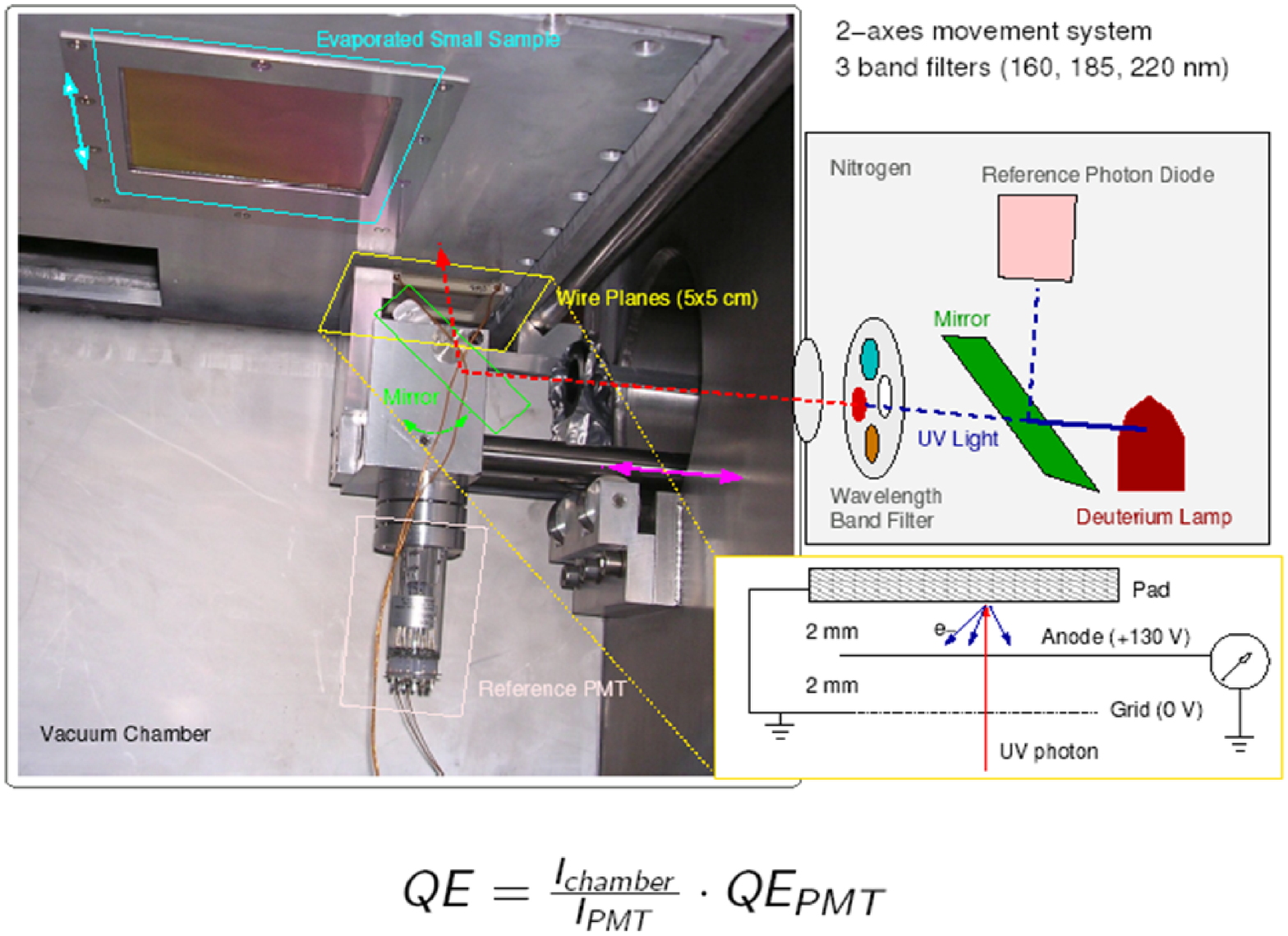}
\caption{The Quantum Efficiency measurement system.}
\label{evap2}
\end{figure}

A dedicated facility has been built for CsI evaporation of large area 
photocathodes. It consists of a cylindrical stainless steel vessel 
(110 cm height, 120 cm in diameter) equipped with four crucibles containing 
CsI powder (see Fig.~\ref{evap1}).  A vacuum of a few $10^-7$ mbar can be 
reached in less than 24 h. The prepolished pad plane (a printed circuit
with three layers of metals, nickel, copper, and gold, glued on the 
vetronite substrate) is housed in the vacuum chamber and heated to $60^\circ$ 
C usually for $12\!-\!24$ h. The location of the crucibles with respect to the 
photocathode and their relative distance are optimized to ensure a minimum
variation in thickness of 10\% using equal amount of CsI in each crucible. 
The CsI powder evaporates at a temperature of $\sim 500$  C. In order to 
monitor the quality of the evaporation and its uniformity, an online 
quantum-efficiency measuring system has been built and successfully 
employed~\cite{cusanno03} (see Fig.~\ref{evap2}). A movement system allows 
%
%
%
%
us to map out the entire 
photocathode. A deuterium lamp has been used as UV source light. The UV 
collimated beam (1 cm in diameter) is split by means of a semi-transparent 
mirror in such a way to allow monitoring of the lamp emission by measuring 
the current from a photodiode. 
Three narrow band filters (25 nm FWHM spread) selecting respectively 160, 
185 and 220 nm have been employed. The UV beam is sent, through a rotatable
mirror, to the photocathode. The photocurrent, generated by electrons 
extracted from the CsI film, is detected with a small ($5\times 5$ cm$^2$) 
wire chamber  located at a distance of 2 mm from the photocathode. The wires 
have a collection voltage of 133 V. A second wire plane, behind the first and
oriented perpendicular to it, is kept at ground potential to obtain good 
charge collection on the first plane. After measuring the wire-chamber
photocurrent (A2), the light is sent to a calibrated PMT, used in diode 
mode (A1), by rotating the mirror. The currents (1250 nA range) are measured
by a picoammeter (KEITHLEY 485). The ratio of the currents A2/A1, multiplied 
by the PMT quantum efficiency, gives the ``absolute'' quantum efficiency 
of the photocathode.
Following the prescription of the ALICE HMPID evaporation system, we have 
operated our system in such a way to deposit a 300 nm CsI film. This 
thickness should guarantee safe operation of photocathode. In fact no 
difference in quantum efficiency has been observed in the $150-700$ nm 
range~\cite{cusanno03}.
The thickness of 300 nm has been chosen as a compromise for having a 
``stable'' photocathode, while avoiding charging up problems at high
radiation fluxes. An evaporation speed of 2 nm/s has been chosen as a 
compromise between the need of avoiding CsI dissociation (high crucible
temperature, high speed) and the need of avoiding residual gas pollution on 
the CsI film surface~\cite{cusanno03}.

%
%
\section{Data analysis}
\label{sec:analysis}

\subsection{Missing Energy reconstruction}
\label{subsec:missing-energy}

Event by event, the values of the missing energy were reconstructed by using
the detected momenta in the HRS arms and the incident beam energy, assuming
the mass of the target nucleus and neglecting its recoil momentum.  
The missing energy is computed from
\begin{equation}
E_{miss} = m_K - M_A + \sqrt{(\omega + M_A - E_K)^2-(\vec{q}-\vec{p_k})^2}\ .
\end{equation}

The central value and the spread of the beam energy were continuously monitored
by OTR or SLI measurements and by the Hall A Tiefenback measurement, 
respectively. Those values were added to the data stream every 30~s. 

\subsection{Event Selection}
\label{subsec:event-selection}

In the selection of the events, significant data reduction is obtained by 
applying track quality selections and the PID requisites on the threshold
 \v{C}herenkov counters, shower counters, and RICH detector. Only events 
in which the particle traveling HRS-L was a kaon and the particle traveling 
HRS-R was an electron were selected. 

In addition, selection on the value of the HRS-L/HRS-R coincidence time 
(2~ns window) were applied to the event in order to be included in the 
calculation of the missing-energy spectrum.
Events corresponding to invalid values of OTR or SLI were excluded.

\subsection{Particle IDentification (PID)}
\label{subsec:particle-identification}

\begin{figure}[t] 
\centering
\includegraphics[width=9cm]{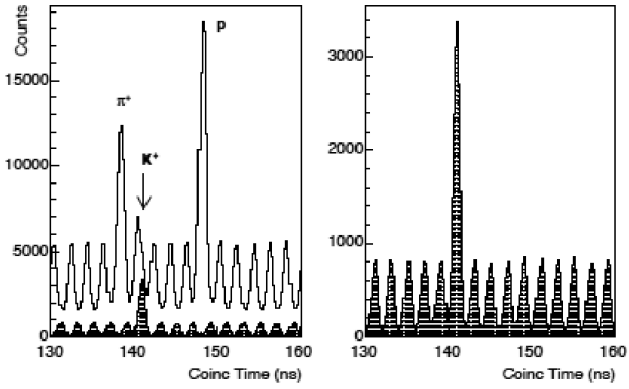}
\caption{Hadron plus electron arm coincidence time spectra. 
Left panel: the unfilled histogram is obtained by selecting kaons with 
only the threshold aerogel \v{C}herenkov detectors. The filled histogram 
(expanded in
the right panel) also includes the RICH kaon selection. The remaining 
contamination is due to accidental $(e,e')\otimes (e,K^+)$ coincidences. 
The $\pi$ and p contamination is clearly reduced to a negligible contribution.}
\label{carb}
\end{figure}

As pointed out previously the PID capability of the HRS's, basically 
guaranteed by TOF, by shower counters in HRS-R and by aerogel counters in 
the HRS-L, is not sufficient for unambigous kaon identification. A RICH was 
built for this purpose. The fundamental role of the RICH in identifying the 
kaons is shown in Fig.~\ref{carb}.

%
%

In the left panel, the unfilled timing spectrum of coincidences between 
the electron and the hadron spectrometers, obtained by selecting for 
kaons using the two threshold aerogel counters, shows a barely visible 
kaon signal with a dominant contribution from mis-identified pions and 
protons. The flat part of this spectrum is given by random coincidences. 
The 2 ns structure is a reflection of the pulse structure of the electron 
beam. The filled spectrum and its exploded version (right panel), is 
obtained by adding the RICH to the kaon selection. Here, all contributions 
from pions and protons completely vanish.

\begin{figure}[t] 
\centering
\includegraphics[width=7cm,clip]{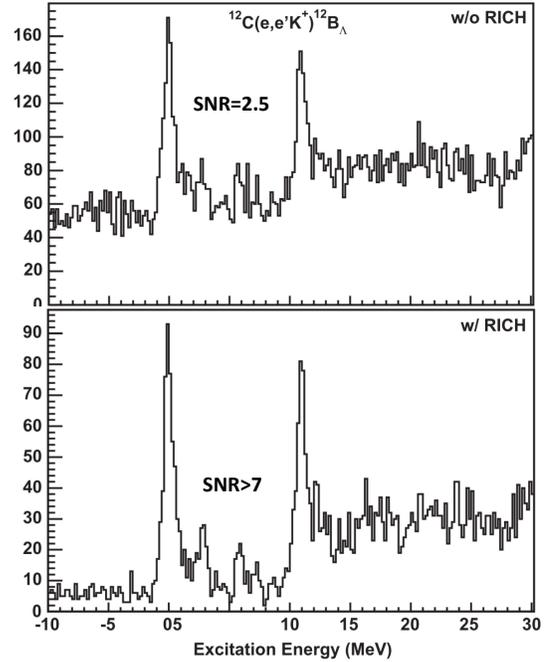}
\caption{Excitation energy spectra of \lam{12}{B} with and without 
the RICH analysis.}
\label{richnorich}
\end{figure}

 The crucial role of the RICH can be seen also from Fig.~\ref{richnorich} 
that clearly shows that the core-excited states of \lam{12}{B} are only barely 
seen if the RICH is not used in the analysis. In that case, the signal to 
noise ratio is insufficient.

%
%

\subsection{RICH}

A new particle rejection algorithm based on the $\chi^2$ test was employed 
with the RICH used in the E94-107 experiment to distinguish kaons from 
pions and  protons. It can be essentially summarized in the following
steps (more details can be found in reference \cite{urciuoli09}).

\begin{itemize}

\item Identification of the minimum ionizing particle (MIP) and 
\v{C}herenkov photon hit points on the RICH cathode. When an MIP
crossed the RICH, and it and the \v{C}herenkov photons generated 
in the RICH radiator hit the RICH cathode plane, the pads near 
their hit points on the cathode generated 
charge signals. In the following, we refer to the single series of 
contiguous cathode pads “fired” by the MIP and the \v{C}herenkov photons on 
the RICH cathode plane as “clusters”. The cluster corresponding to the 
MIP hit point was easily identified by calculating the interception point 
between the particle track provided by the drift chambers located on the 
focal plane of the HRS spectrometer and the RICH pad plane. The maximum 
charge cluster inside a defined radius R around this point was assumed 
to be the one generated by the MIP. All the other clusters on the cathode 
plane whose distance to the MIP cluster position were compatible, within 
the experimental uncertainties, with the generation of a \v{C}herenkov photon 
in the RICH radiator by a proton, kaon or pion whose momentum was equal 
to the one measured by the HRS spectrometer were considered as candidate 
to be generated by a \v{C}herenkov photon hitting the cathode plane. A cluster 
could be made up by one or more pads.

\item Cluster resolving. The presence of two or more relative maximums in the 
geometric distribution in the RICH cathode plane of the pad signals of one 
single cluster indicated that that cluster was produced by two or more 
\v{C}herenkov photons whose hit points on the cathode plane were so close that 
their corresponding clusters geometrically overlapped. These clusters were 
resolved (that is were decomposed into their constituent clusters) by
considering that they were generated by a number of \v{C}herenkov photons equal 
to the cluster relative maximum number. The charge assigned to each of the 
single clusters constituting an unresolved cluster was proportional to the 
charge of the corresponding relative maximum.

\item Single-photon \v{C}erenkov angle determination. The emission angle of 
each single \v{C}herenkov photon generated by the MIP in the RICH radiator was 
determined from the relative position of the corresponding cluster and the 
MIP cluster in the RICH cathode pad plane and from the direction of the 
particle track with respect to the normal to the RICH cathode pad plane, 
using an algorithm based on a geometrical back-tracking.

\item  Particle identification based on the $\chi^2$ test. After the MIP 
cluster identification and the determination of $N$ \v{C}herenkov angles by the 
back-tracking from the $N$ resolved clusters candidate to be generated by 
\v{C}herenkov photons were performed, three $\chi^2$ tests were performed, one 
for each of the three possible hypotheses (proton, kaon or pion) for the 
MIP crossing the RICH. In fact, the measured \v{C}herenkov angle distribution 
around the true value was expected to follow with good approximation a 
Gaussian distribution. As a consequence the sum  $\sum_i \frac{\theta_{expected}
 - \theta_i}{\sigma^2}$, with $\theta_i$ the $i^{th}$ \v{C}herenkov angle 
measurement, $\sigma$ the \v{C}herenkov angle measurement distribution, and 
$\theta_{expected}$ the expected \v{C}herenkov photon emission 
angle according to the particle hypothesis, is expected to follow the 
$\chi^2$ distribution if the particle hypothesis is correct and no cluster 
generated by electronic noise was present. The particle was hence identified 
with the one whose corresponding $\theta_{expected}$ value was such that the 
related $\chi^2$ test provided a result acceptable within a predefined 
confidence level. If none of the three $\chi^2$ tests was acceptable, this 
meant that electronic noise was present and one, two, $\dots M$ terms, starting 
with the largest contributor to the $\chi^2$s, were iteratively removed 
until (at least) one of the three $\chi^2$ values, and hence of the particle 
hypotheses, was compatible with the significance level.   

\item Particle identification based on the single-photon \v{C}herenkov angle 
average calculation. Complementary to the particle identification based 
on the $\chi^2$ test was the traditional identification based on the 
calculation of the average of the $N$ $\theta_i$ measurements. This 
average, when the electronic noise is negligible, is distributed around 
the true value with a standard deviation equal to $\frac{\sigma}{sqrt{N}}$ 
and hence its comparison with the three expected \v{C}herenkov emission angles 
corresponding to the three particle hypotheses is a powerful particle 
identification method.  

\item Particle identification based on the combined use of the $\chi^2$ 
test and of the single photon \v{C}erenkov angle average calculation. The 
$\chi^2$ test is a test on the variance of the $N$ \v{C}herenkov angle 
measurement's Gaussian distribution. The check on the average of the $N$ 
\v{C}herenkov angle measurements is a test on the mean of this distribution. 
Mean and variance of a Gaussian distribution are independent parameters. 
It can be mathematically demonstrated that the $\chi^2$ test and the test 
on the average of the $N$ \v{C}herenkov angle measurements are hence two 
independent tests and can be used simultaneously to obtain proton and 
pion rejection factors nearly equal to the product of the single test 
rejection factors, the deviation from an exact product being due to 
analysis speed considerations and to the presence of electronic noise.
    
\item Use of the aerogel \v{C}erenkov detectors for an independent complete 
PID. The \v{C}herenkov detectors were used in addition to the RICH to obtain 
a 100\% proton and pion rejection with no loss of kaon detection efficiency.

\end{itemize}

 The combined use of the two algorithms provided, in combination with the 
thresholds of two aerogel \v{C}erenkov detectors,  a completely 
satisfactory pion rejection ratio greater than 30000 with practically no 
loss of statistics. 

Based on checks against expected values of the average and the variance 
of the experimental measurements (two statistically independent variables),
this algorithm can be employed not only with the RICH's but 
whenever one deals with detectors that provide independent multiple 
measurements of variables with a constant probability distribution function.

\subsection{Normalization}

In order to calculate absolute cross sections, the missing energy spectrum
has to be properly normalized. The cross section for a level $i$ is computed 
as 
\begin{equation}
  \sigma_i = \frac{N_i}{l~\textrm{surv}(k)\epsilon_e \epsilon_H \epsilon_{coinc} 
\Delta_e \Delta_k \Delta p_e} \ ,
\end{equation}
where $N_i$ is the number of counts in the level i, corrected for the 
deadtime, l is the luminosity, surv(k) is the kaon survival probability 
inside the left arm of HRS, $\epsilon_e$ and $\epsilon_k$ are the detector 
efficiencies for the two HRS arms, $\epsilon_{coinc}$ is the efficiency of 
the coincidence trigger, $\Delta_e$ and $\Delta_k$ are the HRS geometric 
acceptances for the two arms, and $\Delta p_e$ is the momentum acceptance 
for electrons.

Since we consider bound states, $p_k$ and $p_e$ are correlated and the 
cross section is integrated on the full range of $\Delta p_k$. 

The luminosity is controlled by means of beam-current monitors 
and rates of single tracks in HRS arms. The dead time is controlled by 
means of proper data acquisition software. Detector efficiencies are 
controlled by specific analysis software. 

\subsection{Beam Current}

The measurement of beam current is crucial for cross section determination.
For this purpose, the beamline is equipped with two beam-current monitors
about 24.5~m upstream of the target. A beam-current monitor is a 
cylindrical resonant cavity made of stainless steel with a resonant 
frequency matching the frequency of the electron beam. We used the average 
value of the two beam-current monitors for our luminosity calculations. 

\subsection{Single Rates}

Rates of tracks in single HRS arms were continuously monitored in order to
cross check the stability of the luminosity and the proper operation 
of the detectors. If a run periods was showing anomalous values of 
single rates, it was excluded from the cross-section calculation.

\subsection{Efficiency}

We calculated the efficiency of the counter detectors based  on the 
Poisson distribution. Then the efficiency for more than one 
photoelectron detector is $\epsilon = 1 - e^{- N_{p.e.}}$. 

The efficiency of the RICH detector was determined by the usage of 
clean track selection on A1 and A2. For the other components of the
detector package, the standard procedures established for the HRS were 
used~\cite{alcorn04}.

The stability of the detector efficiency was continuously monitored 
for each component of the HRS package. In fact, the track rates 
of the individual detectors were compared to the corresponding luminosity. 

\subsection{Peak Search}

A $\chi^2$-based method was used for the detection of the peaks 
in the missing-energy spectra. This method analyzes energy intervals
in the spectrum and the width of the intervals is variable in a range 
consistent with the energy resolution of the experiment. 
The background in the region of interest is very well reproduced by a 
linear fit. Then for each energy interval showing an excess of counts 
with respect to the background, the confidence level of those counts 
was compared to the fluctuation of the corresponding background. 
If the confidence level was larger than 99\% and a local maximum was 
found, then the corresponding energy region was fitted with a Gaussian 
or Voigt curve. 

\subsection{Energy Resolution}

Since the energy resolution is critical for the experimental results, the 
best computation of all the terms involved in the calculation of the missing 
energy have to be as precise as possible. Therefore:
\begin{itemize}

\item The optics database for both the arms of HRS has to provide 
the best momentum resolution in an acceptance range as large
as possible.
\item The beam-energy spread was continuously monitored using OTR and
SLI in order to exclude the events when the energy spread was not good.
\item The central beam energy was continuously monitored. 
\item In the case of a rastered beam, a software procedure was used to 
evaluate the real position of the incident electrons, to correspondingly
compute the entrance position of the particles in HRS and thus their momentum.
\item An iterative method to check the presence of an unphysical dependence 
of the missing mass on the scattering variables was performed.

\end{itemize}

\subsection{Radiative corrections}
\label{subsec:radcorr}

\begin{figure} 
\centering
\includegraphics[width=6.4cm]{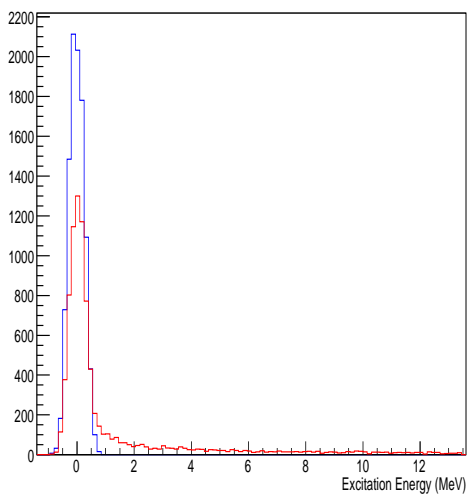}
\caption{Color online. From Ref.~\cite{urciuoli15}. One peak of the excitation 
energy spectrum of the hypernucleus \lamb{9}Li obtained through the reaction 
$^9$Be$(e,e'K^+)$\lamb{9}{Li} as predicted by the Monte Carlo SIMC
when including all effects (red curve) and turning off the
radiative effects (blue curve). Arbitrary units. The position
of the peak has been made coincident with the ground state.}
\label{SIMC-peaks}
\end{figure}

Standard radiative unfolding procedures were performed for 
\lam{12}{B}~\cite{iodice07} and \lam{16}{N}~\cite{cusanno09} hypernuclei while, 
due to the more complicated structure of the spectrum, a different and 
relatively new technique was used for \lamb{9}Li. Here we summarize briefly 
this technique. Details can be found in Ref.~\cite{urciuoli15}.
In the case of \lamb{9}Li we have utilized the property, mathematically 
demonstrated in Appendix A of Ref.~\cite{urciuoli15}, that the subtraction 
of radiative effects from an experimental spectrum does not depend on the 
hypothesis/choice of the peak structure used to fit the spectrum itself, 
providing that the fit is good enough. This property is very useful when 
the peak structure underlying an experimental spectrum is uncertain and
several theoretical (or simply hypothetical) peak structures fit the 
experimental spectrum well and it is not obvious which of these structures is 
``the right one''. The $^9$Be$(e,e'K^+)$\lamb{9}{Li} reaction with the 
E94-107 experimental apparatus was simulated with the Monte Carlo SIMC. 
A single excitation-energy peak produced by this simulation is shown by 
the red curve in Fig.~\ref{SIMC-peaks} (position and amplitude of the 
%
%
peak are arbitrary). The same figure shows, as a blue curve, a single 
excitation-energy peak produced by Monte Carlo SIMC simulations in the same
conditions but with radiative effects ``turned off''. Several peak 
configurations, with different number, position and heights of peaks like 
the one reproduced by the red curve of Fig.~\ref{SIMC-peaks} fit the 
\lamb{9}Li experimental energy spectrum. Because of the properties of 
the subtraction of radiative effects from spectra quoted above, all of 
them produced the same radiative-corrected spectrum determined by turning 
off the radiative corrections in the SIMC simulations, that is by substituting
the Fig.~\ref{SIMC-peaks}-red-curve-like peaks with peaks like the one 
reproduced by the blue curve of Fig.~\ref{SIMC-peaks}. Because the Monte 
Carlo fits to the experimental spectrum were not perfect, slightly different 
radiative corrected spectra were obtained from the different peak 
configurations. The biggest of these differences was assumed as the 
systematic error generated in the reconstruction of the radiative corrected 
spectrum by the method employed to generate it. This systematic error was in 
any case negligible compared to the statistical error. 

The unfolding of radiative corrections has been done bin-by-bin. 
Defining the ``Radiative Corrected Monte Carlo'' spectrum as the 
radiative-corrected spectrum obtained with the procedure described above 
and the ``Regular Monte Carlo'' spectrum as the spectrum produced by the 
SIMC simulations without turning off radiative corrections that fits 
the experimental spectrum (this spectrum could be obtained, as quoted 
above, with different peak configurations), the content of each bin of 
the radiative-corrected spectrum was obtained by multiplying the 
corresponding bin of the experimental spectrum by the correction factor 
given by the ratio of the Radiative Corrected Monte Carlo spectrum and 
the Regular Monte Carlo spectrum for that bin. In order to avoid possible 
removals of background enhancements or to artificially zero the spectrum 
in the regions where the Radiative Corrected Monte Carlo spectrum was 
zero, the ratio between the Radiative Corrected Monte Carlo spectrum 
and the Regular Monte Carlo spectrum was performed after summing
the background for each of them. The background value
was then subtracted from the result of the product of the
ratio with the corresponding bin.

Once the radiative corrections were applied, the binding-energy spectrum 
resolution is small enough to clearly show the three-peak structure shown 
in Fig.~\ref{databe}.

\subsection{Calibrations}

\subsubsection{Optics}

The quality and exact character of the optics transformation tensor were 
measured with a series of elastic scattering measurements using a 2 GeV 
electron beam on C and Ta targets. Measurements were also made using a 
sieve-like mask in front of each spectrometer to optimize and calibrate 
the angular reconstruction. Finally, a check on residual correlations 
between the missing energy and the optics variables was performed by a 
dedicated iterative method. This method~\cite{optics} was based on the 
property that any change in the optical data base corresponds mathematically 
to an addition, to the missing-mass numeric value, of a polynomial in the 
scattering coordinates of the secondary electron and of the produced kaon. 
The method consisted of checking whether the numerical missing-mass value 
produced by the optical data base had unphysical mathematical dependencies  
on the electron and kaon scattering variables. Fitting these mathematical 
dependencies with a polynomial $P$, the method consisted in finding the 
change in the optical data base that produced an addition to the calculated 
numerical value of the missing mass equal to $-P$ and that hence eliminated 
the unphysical missing-mass dependency. Once any possible dependency of 
the numerical value of the missing mass on the scattering coordinates had 
been eliminated with the method described above, the optic data base was 
optimized. In fact, any further change in the optic data base would have 
meant the addition of a polynomial in the scattering coordinates to the 
numerical value of the missing mass that would have produced new unphysical
dependencies. The method described above is based on physics considerations. 
It also usually produces the best resolution. In fact, unphysical dependencies 
of the missing-mass numerical value on the scattering coordinates means that 
the missing-mass values as produced by the optic data base spread around 
the true binding-energy values as function of the scattering coordinates 
increasing the FWHM of the missing-energy spectrum peaks.   
%
%
\begin{figure} 
\centering
\includegraphics[width=9.9cm]{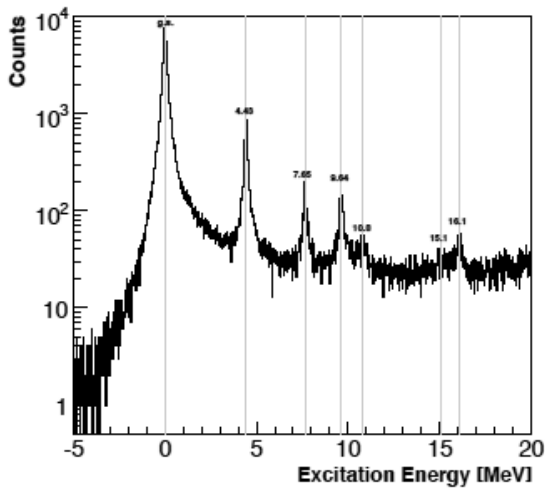}
\caption{Elastic $^{12}$C scattering spectrum as seen in one arm of the 
HRS + Septum configuration after optimization. The width of all the peaks, 
elastic and inelastic, is $10^{-4}$ (FWHM}
\label{optics}
\end{figure}
  
The results of the calibration and optimization effort are illustrated in 
Fig ~\ref{optics}. 

\subsubsection{Waterfall target}

A calibration of the target thickness as a function of pump 
speed has been performed.The thickness was determined from 
the elastic cross section on hydrogen~\cite{garibaldi92}. The target 
thickness used was 
$75\!\pm\!3\,(\mathrm{stat.})\pm\!12\,(\mathrm{syst.})$ mg/cm$^2$.

\subsubsection{Energy scale}

Careful calibration methods were employed to determine the binding-energy 
spectra of the hypernuclei \lam{16}{N} and \lamb{9}{Li}, and of the 
excitation-energy spectrum of the hypernucleus \lam{12}{B}. These methods 
were necessary because the actual kinematics of the processes producing 
the hypernuclei quoted above differed from the nominal ones by amounts 
that would have produced significant shifts and distortions in 
binding-energy and excitation-energy spectra if proper measures 
had not been taken. In fact, while the actual kinematics values in the 
experiment, provided by the CEBAF accelerator electron beam energy and by 
the central momenta and angles of the HRS electron and hadron arms, could 
be considered constant for the entire course of the experiments (their 
variations being of the order of $10^{5}$ for the CEBAF electron-beam energy 
and the central momenta of the HRS electron and hadron arms, and practically 
zero for the spectrometer central angles), they differed by unknown amounts, 
referred to as ``kinematical uncertainties'' in the following, from their 
nominal values, that is the values the CEBAF beam energy and the HRS central 
momenta and angles were nominally set at according to the kinematics of the 
experiment. Although small (the experimental uncertainties on the CEBAF 
accelerator electron-beam energy and on the spectrometer central momenta 
being of the order of $10^{4}$ to $10^{3}$ and those on the spectrometer 
central angles of the order of $10^{2}$), these kinematical uncertainties 
have two non-negligible effects a) they cause global shifts in the 
binding-energy spectra and b) they cause peak distortions increasing their 
FWHM in the binding/excitation-energy spectra. The actual kinematics values 
in an experiment are then those that position states at their known value 
in binding/excitation energy spectra and minimize peak FWHM's.  

%
%
\begin{figure} 
\centering
\includegraphics[width=9cm]{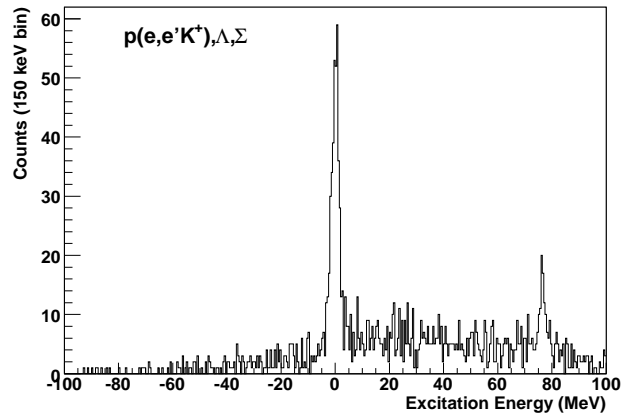}
\caption{Excitation energy spectrum of the $p(e,e'K^+)\Lambda,\Sigma^0$ 
on hydrogen used for energy scale calibration. The fitted positions
(not shown on the plot) for the peaks are $-0.04\pm 0.08$ MeV and
$76.33\pm 0.24$ MeV.}
\label{fig:calib}
\end{figure}

To calibrate the binding-energy scale for \lam{16}{N}, the 
$\Lambda$ peak position from the reaction on hydrogen was first obtained 
using the nominal central values for the  kinematic variables, and then 
constrained to be zero by applying a small shift to the energy of the beam
(the quantity with the largest uncertainty). This shift is common to 
reactions on hydrogen and oxygen and therefore its uncertainty
does not affect the determination of the binding energies of the
\lam{16}{N} levels. A resolution of 800 keV FWHM for the $\Lambda$ peak on
hydrogen is obtained. The linearity of the scale has been verified
from the $\Sigma^0 -\Lambda$ mass difference of 76.9 MeV. For this purpose, 
a few hours of calibration data were taken with a slightly lower
kaon momentum (at fixed angles) to have the $\Lambda$ and $\Sigma^0$
peaks within the detector acceptance. Fig.~\ref{fig:calib} shows the two peaks 
associated with p$(e,e'K^+)\Lambda$ and p$(e,e'K^+)\Sigma^0$ production.
The linearity is verified to $(76.9 - 76.4\pm 0.3)/76.4 = 0.65\pm 0.40$\%

The hypernuclei \lam{12}{B} and \lamb{9}{Li} were produced in 
one run where waterfall or hydrogen targets were not available. 
For these two hypernuclei, the energy-scale calibration was hence performed 
by positioning, in the \lam{12}{B} binding-energy spectrum, the ground-state 
peak at its known value of $-11.37$ MeV determined by emulsion data, after
taking into account the additional shift in the energy scale (calculated 
through Monte Carlo simulations), caused by the energy losses in the target 
$^{12}$C by the participants to the reaction producing the hypernucleus
\lam{12}{B}. The kinematical uncertainties were further reduced by minimizing 
the width of \lam{12}{B} ground-state peak. This peak is actually a doublet 
with its two components separated by $\sim 160$ keV. However, this value 
is small enough with respect to the energy resolution of the experiment to
make the approximation of assuming the \lam{12}{B} ground state as
a single peak still valid and make consequently small the distortions 
incidental to the minimization of the FWHM of a peak that is actually a 
doublet. No attempt to minimize the FWHM was performed on the other peaks 
of the \lam{12}{B} spectrum. Because the hypernuclei \lam{12}{B}
and \lamb{9}{Li} were produced with the same apparatus and the same nominal 
kinematic variables, the \lam{12}{B} excitation-energy calibration results 
where applied to obtain the \lamb{9}{Li} binding-energy spectrum, after 
taking into account the difference between the global shifts of the peaks
in the two spectra due to the difference of the particle-energy loss
in the $^{12}$C and in the $^9$Be targets. 

\subsection{Systematic errors}

The main sources of systematic errors in the missing-energy spectrum are:

\begin{itemize}

\item The uncertainty on the value of the beam energy.
\item The uncertainty on the values of the track momenta.
\item The uncertainty on correction for radiative effects.

\end{itemize}

If not specified, our systematic errors on the position of the peaks 
in the missing-energy spectrum are negligible with respect to their 
corresponding statistical errors.

For the calculation of the binding energies, an additional contribution to
the systematic error has to be considered, due to the need for an absolute 
energy scale. 
 In the case of \lam{12}{B} the binding energies were not calculated.
In the case of \lam{16}{N} this contribution is determined by 
the uncertainty in the position of the $\Lambda$ peak obtained from the 
strangeness production on hydrogen in the waterfall target. 
 In the case of \lamb{9}{Li}, an additional contribution to the 
systematic error is due to the uncertainty of the knowledge of the 
\lam{12}{B} ground-state binding energy, which we used as reference.

For the calculation of absolute cross sections, the following sources of 
systematic uncertainties were considered:

\begin{itemize}

\item The uncertainty on the integrated beam current. 
\item The uncertainty on the target thickness. It is 2\% for solid targets. 
For the oxygen in the waterfall target it is 16\% as previously quoted.
\item The uncertainty on the detector efficiencies.
\item The uncertainty on the dead-time correction 
\item The uncertainty on the HRS phase space.
\item The uncertainty on the corrections for radiative effects.

\end{itemize}

Based on the run-by-run fluctuations, we evaluated our global systematic 
error on absolute cross sections as being within 15\% for \lam{12}{B}
and within 20\% for \lam{16}{N} and \lamb{9}{Li}.
Due to the different contributions of the radiative effects, 
systematic errors were individually calculated for each peak in the 
missing-energy spectra.

%
%
\section{Theory}
\label{theory}
\subsection{Electroproduction of hypernuclei in DWIA}

Production of hypernuclei by a virtual photon associated with 
a kaon in the final state can be satisfactorily described in the 
distorted-wave impulse approximation~\cite{sotona94} because the 
photon and kaon momenta are rather high ($\approx 1$ GeV). 
The cross section for the production of the ground or 
excited states of a hypernucleus depends on the many-particle matrix 
element between the nonrelativistic wave functions of the target 
nucleus ($\Psi_A$) and the final hypernucleus ($\Psi_H$)
\begin{equation}
T_{if}^{\mu} = \langle \Psi_H | 
\sum_{j=1}^Z \chi_\gamma \chi_K^* J_j^\mu | \Psi_A \rangle .
\label{MBME}
\end{equation}
Here $J_j^\mu$ is the hadronic current corresponding to electroproduction 
of a $\Lambda$ on the proton (the elementary production). The sum runs over 
the protons of the target nucleus as we study $K^+$ electroproduction. 
In the one-photon approximation, the virtual photon is described by the 
function $\chi_\gamma$ proportional to the product of the wave functions 
of incoming and outgoing electrons without Coulomb distortion. The kaon 
distorted wave $\chi_K$ is calculated in the eikonal approximation from 
a first-order optical potential in which the density of the hypernucleus 
is approximated by that of the target nucleus. The eikonal approximation 
is sufficient for weakly interacting kaons with momenta larger than 1 GeV. 

The kaon-nucleus optical potential is constructed using the kaon-nucleon 
total cross section and the ratio of the real to imaginary parts of the 
forward scattering KN amplitude. The amplitude is properly isospin 
averaged to take into account the number of protons and neutrons in the 
nucleus. The KN amplitudes for isospin 0 and 1 are calculated in a separable 
model~\cite{bydzovsky99} with partial waves $l=0,1,...7$ and with parameters 
recently fitted to the phase shifts and inelasticity parameters in 
the KN scattering. The nuclear density in the potential is modeled by 
the harmonic-oscillator form with the constant taken from experiments 
on the nucleus charge radii. 

The matrix element is calculated in the frozen-nucleon approximation 
(the target proton three momentum in the laboratory frame is zero) which 
significantly simplifies the integration and allows one to express 
the elementary amplitude in the laboratory frame via only six CGLN 
amplitudes ~\cite{sotona94}. To go beyond this factorization approach, 
i.e. include also a Fermi motion in the nucleus, one would have to 
calculate the elementary amplitude in a general reference frame which 
would, together with the momentum integration, make the calculation 
considerably more complicated.

Experiments on electroproduction of hypernuclei are performed in 
the kinematical region of almost real photons ($Q^2 = -q^2_\gamma\approx 0$). 
In this kinematics, the elementary electroproduction cross section 
is dominated by its transverse part and can be approximated by 
the photoproduction cross section - e.g., as in Ref.~\cite{bydzovsky12}. 
However, even at values of $Q^2$ as small as those in 
Table~\ref{tab:kinematics}, the transverse-longitudinal interference 
contribution can be important. That is why in the calculations presented 
here, the full electroproduction cross section is used~\cite{sotona94}. 
 
\subsection{Elementary production process}

The hadronic current, expressed in the non-relativistic two-component formalism 
via six CGLN amplitudes in the laboratory frame, is calculated using an isobar 
model~\cite{sotona94,skoupil16}.  
Due to the strong damping of the hypernuclear production cross section by 
the nucleus-hypernucleus form factors for large kaon angles, the dominant 
contribution from the elementary amplitudes comes from the region of very 
small kaon angles. In this kinematical region, however, the various isobar 
models give big differences in predicted cross sections, especially for 
$E_\gamma^{lab} > 1.7$~GeV~\cite{bydzovsky12,bydzovsky13,bydzovsky07}, see also 
Sect.~\ref{elem-results}.  
The magnitude of these differences constitutes an important part of the
theoretical uncertainty in predicting the hypernuclear production rate. 
For the energies of the Hall A experiments, $E_\gamma^{lab} = 2.2$~GeV, 
the Saclay-Lyon model~\cite{mizutani98} gives very good results for 
the hypernuclear cross sections~\cite{urciuoli15,iodice07,cusanno09}. 
In our analysis,
we also use a very recent isobar model BS3~\cite{skoupil18} that fits
the new data on photo- and electroproduction well and also gives
reasonable predictions for the cross sections at small kaon angles.
Note that the JLab data on the $Q^2$ dependence of the separated
transverse and longitudinal cross sections are significantly better
described by the BS3 model than by the Saclay-Lyon (SLA) model as
is shown in Fig.~13 of Ref.~\cite{skoupil18}. It is, however, fair to say
that, in contrast to BS3, the SLA model was not fitted to these data.

\subsection{Nucleus and hypernucleus wave functions}

The wave functions for light hypernuclei are obtained from shell-model 
calculations using an effective p-shell interaction to describe the
nuclear core states~\cite{millener08}. In this weak-coupling approach, 
both $\Lambda$ and $\Sigma$ hyperons in s-states are coupled to p-shell 
core wave functions optimized to fit a wide range of p-shell properties.
The $\Lambda$N effective interaction can be written in the form 
\begin{eqnarray}
V_{\Lambda N}(r) = V_0(r) + V_\sigma(r)\vect{s}_\Lambda\cdot \vect{s}_N + 
V_\Lambda(r)\vect{l}_{\Lambda N}\cdot\vect{s}_\Lambda + \nonumber \\
V_N(r)\vect{l}_{\Lambda N}\cdot\vect{s}_N + V_T(r)\;S_{12},
\label{eff-interaction}
\end{eqnarray}  
where $V_0$ is the spin-averaged central interaction, $V_\sigma$ 
is the spin-dependent central term, $V_\Lambda$ and $V_N$ are the 
spin-orbit interactions and $V_T$ is the tensor $\Lambda$N interaction 
with $S_{12} = 3(\vect{\sigma}_{\Lambda} \cdot \vect{r}/r)
(\vect{\sigma}_N \cdot \vect{r}/r) - \vect{\sigma}_{\Lambda} \cdot
\vect{\sigma}_N$. 
The quadratic spin-orbit term, also allowed by symmetries, is 
neglected.

For a p-shell nucleon and a $\Lambda$ in the $s$ orbit 
the radial integrals can be parameterized via five constants, 
$\bar{V}$, $\Delta$, $S_\Lambda$, $S_N$, and $T$
\begin{eqnarray}
V_{\Lambda N} = \bar{V} + \Delta\;\vect{s}_\Lambda\cdot \vect{s}_N + 
S_\Lambda\;\vect{l}_{\Lambda N}\cdot\vect{s}_\Lambda + \nonumber \\
S_N\;\vect{l}_{\Lambda N}\cdot\vect{s}_N + T\;S_{12},
\end{eqnarray}    
which have a one-to-one correspondence with the five  $p_N s_\Lambda$ 
two-body matrix elements. The last four matrix elements can be 
determined from the analysis~\cite{millener08} of precise 
$\gamma$-ray spectra of p-shell hypernuclei obtained via
hypernuclear $\gamma$-ray spectroscopy, mostly with
the Hyperball~\cite{hashtam06}. The $\Sigma N$ and $\Lambda N$-$\Sigma N$ 
coupling matrix elements can be parametrized in the same way with the
values of the parameters calculated using Woods-Saxon wave functions
and Gaussian or Yukawa representations of $YN$ G-matrix elements
based on free $YN$ baryon-baryon potentials~\cite{millener10}.  
The $\Lambda$-$\Sigma$ coupling makes significant contributions
to hypernuclear doublet spacings but, while included in the shell-model
calculations, is not important for analyses of $(e,e'K^+)$ data.

 Unfortunately, $\gamma$-ray spectroscopy is feasible 
only for hypernuclear states lying below particle thresholds. 
Information about the structure of multiplets above 
particle-emission thresholds, generally when the $\Lambda$ is in a 
$p$ orbit, is provided by analyses of the missing-mass spectra from 
electroproduction (reaction spectroscopy) which can be realized with 
better energy resolution than from the pion or $K^-$ induced 
production reactions~\cite{hashtam06}. 

After the partial-wave decomposition of the wave functions 
$\chi_\gamma\chi_K^*\,$,  the many-body matrix element (\ref{MBME}) can be 
expressed by means of the hypernucleus-nucleus structure constants, radial 
integrals, and the CGLN amplitudes. The structure constants are calculated 
from one-body density matrix elements provided by the shell-model structure 
calculations with the interaction (\ref{eff-interaction}). In the radial 
inegrals, we make use of the Woods-Saxon single-particle wave functions for 
the target proton and final $\Lambda$ which we suppose to be a more realistic 
approximation than the harmonic oscillator wave functions, especially 
in the case of weakly bound particles. The parameters, 
the radius, slope, and potential depth of the Woods-Saxon potential, 
which includes the central, spin-orbital, and Coulomb parts,  
are taken from other processes. The single-particle binding energies 
correspond to the particle separation energies. 

 The two-body matrix
elements for hyperons in $0p$ orbits (20 matrix elements for $p_Np_\Lambda$)
for use in the shell-model calculations (with $\Lambda$-$\Sigma$ coupling 
included) are likewise calculated using Woods-Saxon wave functions. 

\subsection{Isobar and Regge-plus-resonance models}

The elementary electroproduction process can be described 
by isobar models based on an effective Lagrangian with 
only hadronic degrees of 
freedom~\cite{mizutani98,KM,williams92,bydzovsky05,skoupil16,skoupil18}. 
Another approach, suited also for description above the resonance 
region up to $E_\gamma^{lab} \approx 16$ GeV, is the Regge-plus-resonance 
model~\cite{decruz12} (RPR) which combines the Regge 
model~\cite{guidal97}, appropriate to description above the resonance 
region ($E_{\gamma}^{lab}>4$ GeV), with elements of the isobar model 
eligible for the low-energy region. 
Both approaches are the one-channel description which neglect interaction 
in the final state violating unitarity. However, they are 
suitable for more complex calculations of electroproduction of 
hypernuclei~\cite{bydzovsky12}. 

Generally, the production amplitude can be split into 
a resonant and nonresonant part. In the isobar and RPR models 
the resonant part is composed of exchanges of nucleon resonances 
in the $s$ channel that can model resonant phenomena in physical 
observables. The nonresonant part in an isobar model consists  
of the Born terms and exchanges of kaon resonances K$^*$ and K$_1$ 
in the $t$ channel and of hyperon resonances in the u-channel. 
In kaon production the contribution from the Born terms is very large 
and reduced assuming either hadronic form factors in 
the baryon-meson-baryon vertices~\cite{KM} or additional exchanges 
of hyperon resonances in the $u$ channel~\cite{mizutani98}. 
In the Gent isobar model a combination of both methods  
is suggested~\cite{janssen01}. The hadronic form factors suppress the Born 
terms very strongly, especially at small kaon angles~\cite{bydzovsky13a}. 
Selection of the method therefore 
considerably influences the dynamics of the isobar model. 
Besides the reduction of the Born terms, the hadronic form factors 
can model an internal structure of hadrons in the strong vertices 
that is neglected in the effective Lagrangian. 

The problem of too large Born contributions is avoided in the RPR 
approach.  In this model, the nonresonant part is composed of  
exchanges of two degenerate $K$ and $K^*$ trajectories. The three free 
parameters can be evaluated in fitting to photoproduction data above 
the resonance region~\cite{decruz12}.  
Note that no hadronic form factors in the nonresonant part 
are needed. The different description of the nonresonant part is the main 
difference between the isobar and RPR models which is important 
for very small kaon angles~\cite{bydzovsky13a,bydzovsky13}, 
see also Sect.~\ref{elem-results}.

%
%
\section{The results}

\subsection{Elementary reaction}
\label{elem-results}

The elementary reaction, the $\Lambda$ production mechanism, is fundamental to 
the interpretation of hypernuclear data~\cite{urciuoli15,iodice07,cusanno09}. 
The reaction has to be studied, especially, at forward kaon angles 
($\theta^{c.m.}_K < 30^\circ$) where there is a lack of data and a wide 
disagreement among existing 
models~\cite{bydzovsky12,bydzovsky13,bydzovsky07}. A realistic 
description of the elementary process at the forward angles is decisive for 
an accurate prediction of hypernuclear excitation spectra~\cite{bydzovsky12}. 
Measurements performed at very small values of the virtual-photon mass 
($Q^2 \approx 0$) are important to the understanding of the process with 
virtual photons, in the framework of an effective Lagrangian this means 
extending of our knowledge about the couplings of the virtual photon 
with baryon fields (the longitudinal 
couplings)~\cite{bydzovsky13,skoupil18,achenbach12}.

\begin{figure} 
\centering
\includegraphics[width=5.5cm,angle=270]{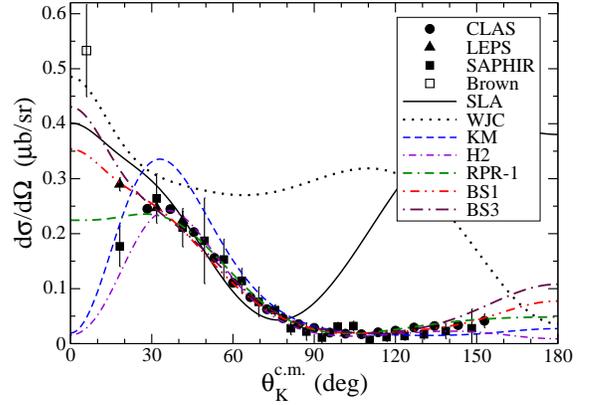}
\caption{Differential cross sections for photoproduction are plotted as 
a function of the center-of-mass angle at $W=2.21$ GeV. Photoproduction data 
are from Refs.~\cite{mccracken10}(CLAS), \cite{sumihama06}(LEPS), and 
\cite{glander04}(SAPHIR). The electroproduction data `Brown'~\cite{brown72} is 
very near to the photoproduction point at $Q^2$ = 0.18 (GeV/c)$^2$ and 
W = 2.17 GeV. The larger angle data can constrain the curves but the 
utility of new small-angle results is evident.} 
\label{elementary1}
\end{figure}

The study of $p(e,e'K^+)Y$ is important not only for the understanding of 
strangeness electroproduction but also for absolute missing-mass calibration 
of the spectrometer systems by using the well known $\Lambda$ and $\Sigma_0$ 
masses. Due to the lack of a neutron target, an absolute mass calibration with 
the hyperon production is impossible for the $(K^-,\pi^-)$ or $(\pi^+,K^+)$ 
reactions. Electroproduction at very forward angles ($\theta^{c.m.}_K < 10^\circ$)
is important to provide reference data for isobar models that give 
inconsistent predictions of the forward-angle cross section, especially at 
the center-of-mass energies $W > 2$ GeV ($E_\gamma^{lab}>1.7$ GeV), 
as shown for photoproduction in Refs.~\cite{bydzovsky12,bydzovsky13}. 
In Fig.~\ref{elementary1} we show the predictions at W = 2.21 GeV of the 
Saclay-Lyon (SLA)~\cite{mizutani98}, Williams-Ji-Cotanch 
(WJC)~\cite{williams92}, 
Kaon-MAID (KM)~\cite{KM}, H2~\cite{bydzovsky05}, recent BS1~\cite{skoupil16}
and  BS3~\cite{skoupil18} isobar models, and of a fit RPR-1~\cite{bydzovsky13} 
to recent data using the Regge-plus-resonance formalism by the Ghent 
group~\cite{decruz12}. The elementary reaction has been studied during the 
E94-107 experiment, using a cryogenic target~\cite{markowitz10}.
%
%

One goal of the current measurement was to determine the angular 
dependence of $d\sigma / d\Omega_K$ at very small angles. Photoproduction 
data from CLAS~\cite{bradford06,mccracken10}, SAPHIR~\cite{glander04}, and 
LEPS~\cite{sumihama06} precisely constrain production models at larger angles 
($\theta^{c.m.}_K > 30^\circ$), e.g., BS1, BS3, H2, and RPR-1 in 
Fig.~\ref{elementary1} fitted to the CLAS data. However, available 
isobar~\cite{mizutani98,williams92,KM,bydzovsky05,skoupil16,skoupil18} and 
Regge-plus-resonance~\cite{decruz12,bydzovsky13} models vary widely 
at $\theta^{c.m.}_K < 30^\circ$, Fig.~\ref{elementary1}, and the previous 
data are not adequate to choose between them. The current measurement provides 
data to constrain the angular dependence as $\theta_K$ goes to zero. 
A second goal was to measure the low $Q^2$ dependence of  $d\sigma / d\Omega_K$ 
to determine a transition from photoproduction with real photons ($Q^2$=0) 
to the photoproduction induced by virtual 
photons~\cite{bydzovsky13,achenbach12}. 
The cross section at low $Q^2$ is also important for studies that 
want to extract the kaon form factor since we can compare to 
extrapolated measurements of the kaon charge radius.

Further, the data determines the $Q^2$ dependence of the 
$\Sigma_0/\Lambda$ production ratio. This ratio drops off rapidly. 
In hadronic production, the ratio decreases from 10 at low energy to 
3 by 1 GeV of energy transfer. In photoproduction, the ratio drops from 2 
at 90 degrees to 0.7 at 22 degrees but data at forward angles 
has not been available. 
In electroproduction, the ratio for the transverse cross section 
drops from 0.7 at the photoproduction point to 0.1 at 
$Q^2 = 1 - 2$ (GeV/c)$^2$, but the behavior in between has not been 
determined. The longitudinal ratio is similar in magnitude to 
the transverse at nonzero $Q^2$. Whether this behavior is just due to 
isospin dependence in the $\Sigma_0$ and $\Lambda$ couplings to 
resonances has not been known.   

The E94-107 kinematics used beam energies of 4.016, 3.777, and 3.656
GeV. The corresponding electron momenta were P$_e$= 1.80, 1.57, 1.44
GeV/c. The kaon momenta was centered P$_K$ = 1.96 GeV/c for the hypernuclear 
running. For these hydrogen measurements, the beam energy was 3.777 GeV, 
P$_e$ was 1.57 GeV/c, and P$_K$ was 1.92, 1.96 and 2.0 GeV/c. The three kaon 
settings both enabled us to slightly extend the range of kinematics 
as well as move the missing-mass peak across the acceptance in a study of 
our understanding. These settings correspond to central values of W = 2.2 GeV 
and $Q^2$ = 0.07 (GeV/c)$^2$. This measurement used currents up of $60 \mu_A$ 
on a 4 cm liquid hydrogen target. The analyzed data was compared to the 
standard Hall A Monte Carlo code, modified to incorporate the septum magnets. 
The comparison to simulation was used to determine the acceptances and put 
cuts on the data to restrict the acceptance to a region where agreement 
between the shapes was excellent between simulation and acceptance.

\begin{figure}[t] 
\includegraphics[width=4.5cm,angle=270]{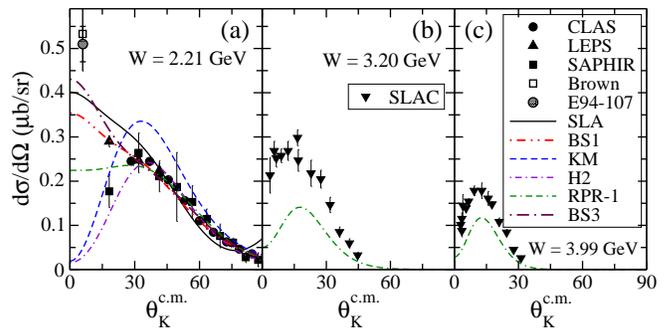}
\caption{The same as Fig.~\ref{elementary1} but for three values 
of the center-of-mass energy. Result of this experiment (E94-107) 
is shown in part (a). 
Predictions of the Regge-plus-resonance model (RPR-1)~\cite{bydzovsky13} 
are comparred with SLAC data~\cite{boyarski69} above the resonance region. 
The problem of normalization of the SLAC data is apparent.} 
\label{elementary2}
\end{figure}

The results are shown in Fig.~\ref{elementary2}(a). Plotted are the 
electroproduction results superimposed on the photoproduction data. Also shown 
are predictions for photoproduction of several models. As can be seen results 
of the models markedly differ for kaon angles smaller than 30$^\circ$. The 
relevant difference in dynamics of the presented models is in their 
description of the nonresonant part of the amplitude. The SLA isobar model 
does not assume any hadronic form factors but instead includes exchanges of 
hyperon resonances to supress contributions form the Born terms, see also 
Sect.~\ref{theory}. The model KM assumes the hadronic form factors without 
any hyperon resonances and the H2, BS1, and BS3 models include both hyperon 
resonances and hadronic form factors. The strong supression of the nonresonant 
part at very small angles is apparent when the hadronic form factors are used 
with or without a small number of hyperon resonances, as in the H2 and KM 
models, respectively. On the contrary, in the recent isobar models BS1 and 
BS3, ample 
number of hyperon resonances with spin 1/2 and 3/2 contribute to the 
nonresonant part of the amplitude which results in a similar behaviour of 
the cross section at $\theta^{c.m.}_K < 30^\circ$ as with the SLA model, 
Fig.~\ref{elementary2}(a). In the Regge-plus-resonance model RPR-1 
the nonresonant part is given by the Regge trajectories without any hadronic 
form factors. The new results are therefore vital 
for understanding the dynamics of models at very small $\theta_K$.
%
%

However, since the data (although at a low $Q^2$) is electroproduction 
not photoproduction, it is possible that the longitudinal amplitudes 
might strongly contribute to the cross section. We estimated the maximum 
contribution using available data on the longitudinal-transverse separations 
of the kaon cross sections. Independent of $Q^2$ and $W$, the data suggest 
a value of $\sigma_L / \sigma_T \approx 0.5$. For this experiment's kinematics, 
this would mean $\sigma_T\approx 0.38$ $\mu$b/sr. This value, that  
corresponds to the photoproduction cross section, rules out 
the models that predict a strong reduction of the cross section at small 
angles, e.g. KM and H2 in Fig.~\ref{elementary2}(a), and favour a steep angular 
dependence for near zero angles predicted by the isobar SLA, BS1, and BS3 
models and by the Regge models (see also Fig. 6 in Ref.~\cite{bydzovsky13a}). 
Note that the SLA model gives the best predictions for the hypernucleus 
excitation functions~\cite{urciuoli15,iodice07,cusanno09} which implies that 
this model provides a realistic description of the elementary process 
at the very small angles that dominate hypernuclear production. 
The forward peaking of the cross section is also consistent with conclusions 
from the analysis of CLAS data~\cite{bradford06}. The authors concluded that 
in the energy region $2.3 < W < 2.6$ GeV,  2.6 GeV being the maximum energy 
in the experiment, the cross section is dominantly forward peaked from
which it can be inferred that a substantial contribution to 
the reaction mechanism comes from $t$-channel exchange.     
The Regge-plus-resonance model, RPR-1, predicts a plateau at small 
angles and energies about 2.2 GeV [Fig.~\ref{elementary2}(a)] showing 
that the Regge-based modeling of the nonresonant part of amplitude can also 
provide reasonable results in this kinematics. 

In Fig.~\ref{elementary2}(b) and (c) we show angular dependence above the 
resonance region at $W$ = 3.20 GeV ($E_\gamma^{lab}$ = 5 GeV) and 3.99 GeV 
($E_\gamma^{lab}$ = 8 GeV), respectively. The SLAC data~\cite{boyarski69} 
and predictions of the RPR-1 model are shown. First note a problem with the 
normalization of the SLAC data~\cite{dey06} which, we suppose, does not 
affect too much their angular dependence. 
In the higher energy region, $W > 3.2$ GeV, the SLAC data  
reveal rather an inverse angular dependence to that observed 
in the resonance region at $W = 2.21$ GeV by this measurement 
and by Bradford \etal\ in Ref.~\cite{bradford06}. 
Therefore, the SLAC data, if their angular dependence does not  
change too much in a re-analysis due to the normalization, 
suggest that the RPR-1 model gives a correct angular dependence 
at very small kaon angles over a large energy region which means  
that at 2.2 GeV a flat angular dependence (plateau) is a more 
realistic behaviour of the cross section. 
Note that some Regge-based models can predict also forward peaking 
cross sections in agreement with the present data but in disagreement 
with the SLAC data, see Fig. 6 in Ref.~\cite{bydzovsky13a}.   
It is obvious that new good quality experimental data for 
kaon c.m. angles 0 -- 20$^\circ$ and in a broader energy region 
are needed to better understand the reaction mechanism.  

The data was also re-binned in three $Q^2$ bins to determine 
the $Q^2$ slope. What is observed is that the differential cross 
section for $\Lambda$ production drops with increasing $Q^2$, while 
the differential cross section for $\Sigma^0$ production is flat. 
A similar re-binning into three $\theta^{c.m.}_K$ bins is 
essentially flat for both $\Lambda$ and $\Sigma^0$, ruling out 
any sharp drop with angle for the $\Lambda$ production. 
The $W$ re-binning data is also flat with energy, as expected 
from photoproduction at larger angles.   
The extracted $\Sigma^0 / \Lambda$ ratio is approximattely 0.5 
and flat with respect to the kinematics. Interestingly, 
this is similar to what the photoproduction data would give, 
extrapolated by a straight line.

%
%
\subsection{Hypernuclear electroproduction}

%
\begin{table*}[th]
\caption{Excitation energies, widths, and cross sections obtained by
fitting the $^{9}$Be$(e,e'K^+)$\lam{9}{Li} spectrum (first three columns)
compared with theoretical predictions using the SLA and BS3 models for 
the elementary interaction(next six columns). A summed 
cross section is given for each of the three doublets 
to compare with the experimental results in the third column.
\label{tab:results}}
\begin{ruledtabular}
\begin{tabular}{ccccccccc}
\multicolumn{3}{c}{Experimental data}  &
\multicolumn{6}{c}{Theoretical cross sections for the SLA and BS3 models} \\
$E_x$ &  Width  &  Cross section & $E_x$  & $J^\pi$ & 
\multicolumn{2}{c}{SLA} & \multicolumn{2}{c}{BS3} \\
 (MeV)  &  (FWHM, MeV) & $(nb/sr^2/GeV)$ & (MeV) &  &
 $(nb/sr^2/GeV)$ & Sum & $(nb/sr^2/GeV)$  &  Sum\\
 \hline
0.00 $\pm$ 0.08  &  0.73 $\pm$ 0.06  &  0.59 $\pm$ 0.15  & 
                  0.00  & $3/2^+$  &  0.164  &  & 0.157 &   \\
0.57 $\pm$ 0.12  &  0.73 $\pm$ 0.06  &  0.83 $\pm$ 0.13  & 
                  0.56  & $5/2^+$  &  1.118  & 1.28 & 1.035 & 1.19 \\
 & & & & & & & & \\
1.47 $\pm$ 0.09  &  0.73 $\pm$ 0.06  &  0.79 $\pm$ 0.07  & 
                  1.42  & $1/2^+$  &  0.353   &  & 0.294 & \\
 & & &            1.45  & $3/2^+$  &  0.327   & 0.68 & 0.343 & 0.64\\
 & & & & & & & & \\
2.27 $\pm$ 0.09  &  0.73 $\pm$ 0.06 &   0.54 $\pm$ 0.06  & 
                  2.27  & $5/2^+$  &  0.130   &  & 0.109 & \\
 & & &            2.73  & $7/2^+$  &  0.324   & 0.45 & 0.315 & 0.42 \\
\end{tabular}
\end{ruledtabular}
\end{table*}

Results from the experiment E94-107 on hypernuclear
electroproduction have been already published and briefly
discussed in Refs.~\cite{iodice07} (\lam{12}{B}),
\cite{cusanno09} (\lam{16}{N}), and \cite{urciuoli15} (\lamb{9}{Li}).
Here, we present new radiative corrected results for
\lamb{12}{B}, similar to what was done for \lamb{9}{Li}.
The experimental results for all targets are compared here with
new theoretical predictions based on improved reaction calculations in DWIA.
The improvement consists mainly in using new structure calculations
for the one-body density matrix elements, corrected kaon distortion,
and taking into account hypernuclear-recoil effects. The latter consists 
in correcting the hypernuclear mass for the excitation energy which 
appears to have a considerable effect on the hypernuclear kinetic energy 
and, especially, for the production cross sections (a few per cent). 
In our previous calculations this was included only in the case of 
the oxygen target~\cite{cusanno09}. We also took
care of a more realistic description of single-particle states
of the initial proton and final $\Lambda$ described by 
Woods-Saxon wave functions. In comparison with our previous
calculations in Refs.~\cite{urciuoli15,iodice07,cusanno09}, we give here 
also results with the new isobar model BS3~\cite{skoupil18}. The 
calculations are also compared with the data from the Hall C experiments 
E01-011 and E05-115 and discussed in subsection~\ref{comparison_AC}.

\subsubsection{The $^{9}$Be target}

There are still some unresolved problems in the spectroscopy of hypernuclei 
in the lower part of p-shell. The spectra of \lam{10}{B} (ground-state doublet 
splitting) and \lam{11}{B} (energy of the $1/2^+$ member of the 
first-excited doublet) as measured in precise $(K^-,\pi^-\gamma)$ and 
$(\pi^+, K^+\gamma)$ experiments are inconsistent with the standard 
shell-model description of p-shell hypernuclei~\cite{millener10}. 
The electroproduction of \lamb{9}{Li} from a $^{9}$Be target can 
hopefully shed new light on this problem. In this case the ground-state
doublet and two excited doublets of \lamb{9}{Li} (all lying below the strong 
neutron-decay threshold) are produced with comparable cross sections. 
In addition, splitting of the ground-state  doublet and the second 
excited-state doublet are predicted to be large enough to be detected 
($\sim 500$ keV), while the first excited-state doublet is predicted to 
be almost degenerate. Fig.~1 of Ref.~\cite{urciuoli15} shows the
detailed shell-model predictions. Note that most of the proton removal
strength for $^9$Be is contained in the first three states of the
$^8$Li core. In addition, the $p_\Lambda$ orbit is unbound at $A\!=\!9$ 
and there is no evidence for sharp $p_\Lambda$ states in the $(e,e'K^+)$
spectrum~\cite{urciuoli15}.

%
%
\begin{figure} 
\centering
\includegraphics[width=6.2cm,angle=270]{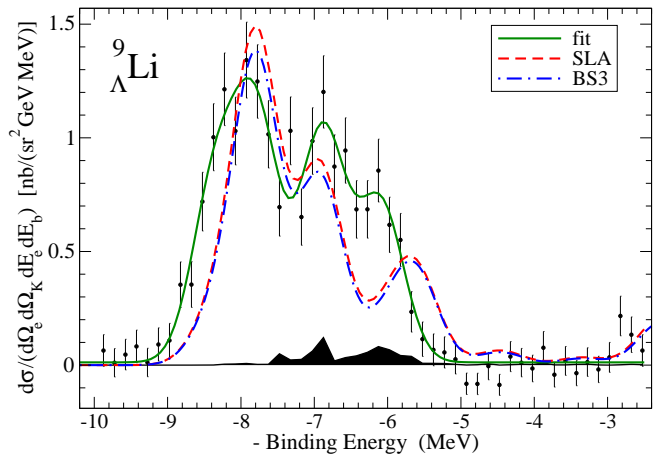}
\caption{The radiative-corrected experimental spectrum of \lamb{9}{Li} in 
comparison with the theoretical prediction (dashed and dash-dotted lines). 
The solid line shows the result of fitting the data with four Gaussians of 
a common width. The theoretical curves were calculated with the width 
extracted from the fit (FWHM = 730 keV).}
\label{databe}
\end{figure}

Figure~\ref{databe} shows the binding-energy spectrum for \lamb{9}{Li} 
production and gives the radiative-corrected experimental data (the points 
with statistical errors)~\cite{urciuoli15} vs. new theoretical results 
(dashed and dash-dotted lines). The band at the bottom of the histogram 
shows the systematic errors. A more detailed description of the procedure 
employed to determine the radiative corrected spectrum can be found in 
Sect.~\ref{subsec:radcorr} and in Appendix A of Ref.~\cite{urciuoli15}.

Once radiative corrections are applied, the binding-energy resolution 
is small enough to clearly show a three-peak structure of the spectrum based 
on the lowest three states of $^8$Li. The experimental spectrum in 
Fig.~\ref{databe} was fitted assuming two Gaussians for the ground-state 
doublet and two Gaussians for the second and third multiplets (solid line). 
The Gaussians were taken to have a common width which was determined to be 
FWHM= 730 keV. A constant background was found negligible in the fit being 
0.2\% at maximum and the $\chi^2_{n.d.f}$ was 1.04. The theoretical curves 
were obtained by superposing Gaussians with an energy resolution of 
730~keV (FWHM). 

The cross sections were calculated using the complete p-shell basis for 
the core nucleus but with a slightly different interaction for the
p-shell core from Ref.~\cite{urciuoli15} which results, e.g. in the 
interchange of the second closely-separated $1/2^+$-$3/2^+$ doublet but 
otherwise negligible changes. 
Moreover, we used realistic Woods-Saxon wave functions for the initial proton 
in the $p_{3/2}$ state bound by 16.89 MeV and the final $\Lambda$ in the 
$s_{1/2}$ state bound by 8.53 MeV.  
Parameters of the kaon distortion were revised utilizing the separable model 
for KN scattering. The hypernuclear recoil was properly included even if 
in this case its effect is not so big as in the case of the other targets. 
The elementary reaction, $p(e,e^\prime K^+)\Lambda$ was described using the 
Saclay-Lyon (SLA)~\cite{mizutani98} and BS3~\cite{skoupil18} models.

The energies, widths, and cross sections extracted from the four-peak fit are 
reported in Table~\ref{tab:results} where they are compared with the 
calculated results for the six lowest states of \lamb{9}{Li}.
The plot in Fig.~\ref{databe} and Table ~\ref{tab:results} show some 
disagreement between the DWIA calculation with a standard model of p-shell 
hypernculei and the measurements, both for the position of the peaks and for 
the cross sections. Specifically, the position of the third multiplet is 
predicted above the value extracted from the data.
The predicted theoretical cross sections are generally in better agreement 
with data than in Ref.~\cite{urciuoli15} but they are still systematically 
10--20\% below the experimental values which we attribute mainly to 
uncertainty in the elementary-production operator~\cite{bydzovsky12}.
Note that the hypernuclear cross sections calculated with BS3 are, in 
general, smaller that those calculated with SLA contrary to a naive 
expectation from a comparison of the elementary cross sections in 
Fig.~\ref{elementary2}(a), where BS3 predicts larger values at 
$\theta_k < 10^\circ$ than SLA. This effect is due to a steeper descent 
of the transverse component as a function of $Q^2$ for BS3 compared with SLA.

\subsubsection{The $^{12}$C target}

$^{12}$C targets have been extensively used in hypernuclear studies
using $(K^-,\pi^-)$, $(\pi^+, K^+)$, and $(K^-_\mathrm{stop},\pi^-)$
reactions dominated by non-spin-flip contributions. 
In the early experiments, only two peaks, separated by about 11 MeV 
and attributed to the $\Lambda$ being in $s$ or $p$ orbit coupled to 
the $^{11}$C ground state, were evident~\cite{hashtam06}. The first 
evidence of structure between the main peaks came from $(\pi^+, K^+)$ 
studies with the SKS spectrometer at KEK 
(E140a, E336, and E369)~\cite{hashtam06}, the best 
resolution of 1.45 MeV being obtained in KEK E369~\cite{hotchi01}. 
Finally, in the stopped $K^-$ experiment of the FINUDA 
collaboration~\cite{agnello05}, further evidence for structrure in this region 
has been observed. However, either because of relatively poor energy 
resolution or statistics, the extraction of energies and cross sections 
from peak fitting was difficult. 
The first electroproduction experiment performed on a $^{12}$C target 
in Hall C~\cite{miyoshi03} had limited statistics but proved that the 
electroproduction process can be used to study hypernuclear spectra with 
a sub-MeV energy resolution and measured cross sections. Further measurements 
in Hall C~\cite{yuan06,tang14} show that a rich structure in the 
$\Lambda$-binding energy spectrum of \lam{12}{B} can be observed with 
a very good energy resolution and that hypernuclear reaction spectroscopy 
is possible.

\begin{figure}[t] 
\centering
\includegraphics[width=8.0cm]{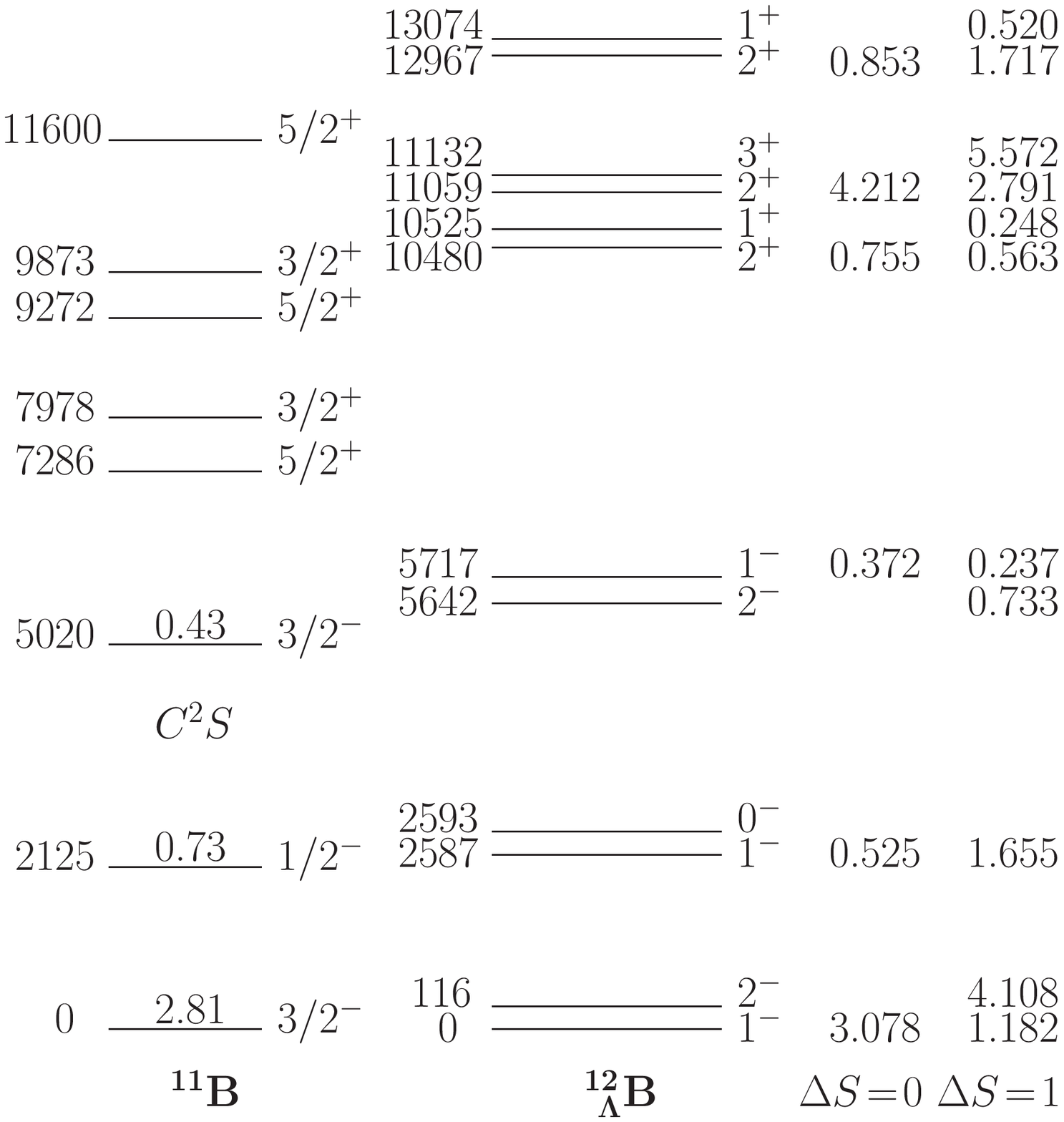}
\caption{The calculated spectrum of \lam{12}{B}. The $^{11}$B core
states are shown on the left along with the spectroscopic factors for 
proton removal from $^{12}$C. All excitation energies are in keV. On the
right, the factors giving the relative population of purely non-spin-flip
($\Delta S\!=\!0$) and purely spin-flip ($\Delta S\!=\!1$) production
reactions on $^{12}$C with $\Delta L\!=\!1$ ($\pi\!=\!-$) or 
$\Delta L\!=\!2$ ($\pi\!=\!+$) are given.}
\label{lb12}
\end{figure}

 The theoretical spectrum for $p^7s_{\Lambda}$ and $p^7p_{\Lambda}$ states
of \lam{12}{B} using the Cohen and Kurath (8-16)2BME 
interaction~\cite{cohen65} for the $^{11}$B core states is shown in 
Fig.~\ref{lb12}. The standard $ps_\Lambda$ parameters for the heavier
$p$-shell hypernuclei from Eq.(4) of Ref.~\cite{millener10} were
used along with $pp_\Lambda$ matrix elements calculated from the
fit-djm potential~\cite{millener10} using Woods-Saxon wave functions
with a binding energy of 0.4 MeV for the loosely-bound $p_{\Lambda}$
orbits. The experimentally-known states of interest for the $^{11}$B
core are shown on the left and stucture factors for non-spin-flip
and spin-flip transitions on the right. The latter give the relative 
population of states for the purely transverse spin operator in
the $(e,e'K^+)$ reaction.

%
%
\begin{figure}[th] 
\centering
\includegraphics[width=6.4cm,angle=270]{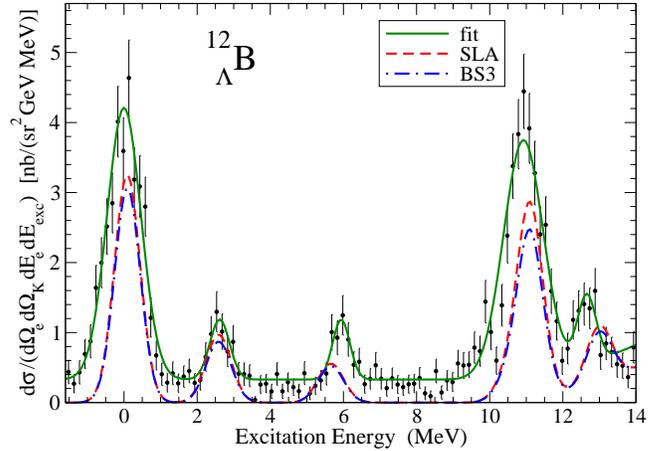}
\caption{The radiative-corrected experimental excitation-energy spectrum of 
\lam{12}{B} (the points with statistical errors) in comparison with the 
theoretical prediction (dashed and dash-dotted lines). The solid line shows 
the result of fits to the data with six Gaussians with independent widths. 
The theoretical curves were calculated with the average width extracted 
from the fit (FWHM = 820 keV) without any background.}
\label{carb3}
\end{figure}

 The splitting of the ground-state doublet in \lam{12}{C} is known
to be 161.5 keV from hypernuclear $\gamma$-ray spectroscopy while
the excitation energies of the excited $1^-$ states are 2832 keV
and 6050 keV~\cite{hosomi15}. The energies of the excited $1^-$
states should be a little higher in \lam{12}{B} and it is clear
that the theoretically predicted excitation energies are more than
300 keV too low.

%
\begin{table*}[t]
\caption{Excitation energies, widths, and cross sections obtained by
fitting the $^{12}$C$(e,e'K^+)$\lam{12}{B} spectrum (first three columns)
compared with theoretical predictions using the SLA and BS3 models for 
the elementary interaction(next six columns). 
\label{tab:Cresults}}
\begin{ruledtabular}
\begin{tabular}{ccccccccc}
\multicolumn{3}{c}{Experimental data}  &
\multicolumn{6}{c}{Theoretical cross sections for the SLA and BS3 models} \\
$E_x$ &  Width  &  Cross section & $E_x$  & $J^\pi$ & 
\multicolumn{2}{c}{SLA} & \multicolumn{2}{c}{BS3} \\
(MeV)  &  (FWHM, MeV) & $(nb/sr^2/GeV)$ & (MeV) &  &
$(nb/sr^2/GeV)$ & Sum & $(nb/sr^2/GeV)$  &  Sum \\
 \hline
0.00 $\pm$ 0.03  &  1.09 $\pm$ 0.05  &  4.51 $\pm$ 0.23 $\pm$ 0.67 & 
         0.00   &  $1^-$   &  0.640  &  & 0.524 &\\
 & & &   0.116  &  $2^-$   &  2.227  & 2.87 & 2.172  & 2.70\\
 & & & & & & & &\\
2.62 $\pm$ 0.06  &  0.64 $\pm$ 0.11  &  0.58 $\pm$ 0.10 $\pm$ 0.11 & 
         2.587  &  $1^-$   &  0.846  &  & 0.689 & \\
 & & &   2.593  &  $0^-$   &  0.001  & 0.85 & 0.071  & 0.76\\
 & & & & & & & &\\
5.94 $\pm$ 0.06  &  0.56 $\pm$ 0.10  &  0.51 $\pm$ 0.09 $\pm$ 0.09 & 
         5.642  &  $2^-$   &  0.368  &  & 0.359 & \\
 & & &   5.717  &  $1^-$   &  0.119  & 0.49 & 0.097 & 0.46 \\
 & & & & & & & &\\
10.93 $\pm$ 0.04 &  1.29 $\pm$ 0.07  &  4.68 $\pm$ 0.24 $\pm$ 0.60 & 
         10.480  &  $2^+$  &  0.194  &  & 0.157 & \\
 & & &   10.525  &  $1^+$  &  0.085  &  & 0.100 & \\
 & & &   11.059  &  $2^+$  &  0.959  &  & 0.778 & \\
 & & &   11.132  &  $3^+$  &  1.485  &  & 1.324 & \\
 & & &   11.674  &  $1^+$  &  0.050  & 2.77  & 0.047 & 2.41\\  
 & & & & & & & &\\
12.65 $\pm$ 0.06 &  0.60 $\pm$ 0.11  &  0.63 $\pm$ 0.12 $\pm$ 0.15 & 
         12.967  &  $2^+$  &  0.552  &  & 0.447 & \\
 & & &   13.074  &  $1^+$  &  0.167  & 0.72 & 0.196 & 0.64\\
\end{tabular}
\end{ruledtabular}
\end{table*}

 The $p_\Lambda$ part of the spectrum should be dominated by the
$2^+$/$3^+$ doublet near 11 MeV in Fig.~\ref{lb12} in
electroproduction. The $p_\Lambda$ doublets are characterized by
the coupling of the $\Lambda$ spin to ${\cal \bm{L}}$ arising from 
the coupling of the core spin to the orbital angular momentum of 
the $p_\Lambda$~\cite{auerbach83}. Two $0^+$/$1^+$ doublets, that 
complete the multiplets of states built on the lowest $3/2^-$ and 
$1/2^-$ of $^{11}$B, are not shown in Fig.~\ref{lb12}. The lower one, 
with the $0^+$ state at 11.197 MeV and the $1^+$ state at 11.674 MeV, 
contains states that should be strongly excited by $\Delta L\!=\!0$ 
transitions from $^{12}$C; the $0^+$ state dominates the $(K^-,\pi^-)$ 
spectrum at small angles~\cite{hashtam06}. The $2^+$ states are excited in
$(\pi^+,K^+)$ reactions while the $2^+$ states and the $0^+$ state
are excited in the $(K^-_{\textrm{stop}},\pi^-)$ reaction.

Results of a new analysis of data from Hall A measurements~\cite{iodice07} 
are presented in Fig.~\ref{carb3} that shows the radiatively unfolded 
excitation-energy spectrum for \lam{12}{B} as  was done in 
the case of \lamb{9}{Li} (the points with statistical errors). 
The spectrum was fitted assuming six Gaussians for the apparent structures 
(multiplets) with independent widths. The background was considered to be 
constant up to the $\Lambda$ separation energy 11.37 MeV and above this 
energy a continuous continuation with a quadratic polynomial was used 
to mimic quasi-free production processes. A very good fit was obtained 
(solid line in Fig.~\ref{carb3}) with $\chi_{n.d.f.}^2=1.05$. 
The widths (FWHM) were obtained in the range of 650--1010 keV where the widths 
of the two main peaks are similar at 990 and 1010 keV. A small peak at an 
excitation energy 9.59 MeV was added due to an apparent shoulder in this 
energy region. However, in comparison with Ref.~\cite{iodice07} 
the radiative corrected spectrum does not support the existence of 
a peak in this region. The origin of the excitation-energy scale was set to 
the peak value of the ground-state (g.s.) level (the uncertainty of the 
absolute scale being about 0.5 MeV). 
The energies, widths, and cross sections extracted from the six-peak
fit  are reported in  Table~\ref{tab:Cresults} where they are
compared with the calculated results for lowest states of \lam{12}{B}.
The theoretical curves in Fig.~\ref{carb3} (dashed and dash-dotted lines) 
were obtained using Gaussians with a width of 820 keV (FWHM) for the energy 
resolution. This width is consistent with values extracted from the fit.

The cross sections in Table~\ref{tab:Cresults} were calculated using new 
nucleus-hypernucleus structure constants for the $\Lambda$ p-state part of 
the spectrum calculated using one-body density-matrix elements from a new 
shell-model calculation including all $p_\Lambda$ and $p_\Sigma$ states. 
In comparison with the previous calculations in 
Ref.~\cite{iodice07}, the kaon-nucleus optical potential was improved and 
the momentum transfer icluded the correction for hypernucleus excitation
energy.  Realistic Woods-Saxon
wave functions for the radial part of the proton and $\Lambda$ wave functions 
were used also for the $\Lambda$ s-state part of the spectrum. The proton in 
$p_{3/2}$ and $p_{1/2}$ states was taken to be bound by 15.96 and 10.37 MeV, 
respectively. The $\Lambda$ in the s-state was bound by 11.37 MeV 
and the p-wave one only by 0.4 MeV. The elementary production was described 
by SLA as in Ref.~\cite{iodice07} and by BS3~\cite{skoupil18}. The 
comparison with the data shows that theory mostly underpredicts the cross 
sections by 20--40\%, similarly to what 
it does in the case of \lamb{9}{Li}. Smaller cross sections in comparison 
with our previous theoretical results in Ref.~\cite{iodice07} are due to a 
stronger kaon distortion which, in general, makes the cross sections smaller. 

Five peaks are observed in the spectrum of \lam{12}{B}, the main ones being 
the g.s. peak and the  p-shell peak at 10.93 MeV. The narrrowest width of 
560 keV has been obtained for the peak at $E_x\!=\!5.94$ MeV whereas the two 
main peaks have widths of about 1 MeV indicating that the experimental 
excitation-energy resolution can be still regarded to be below 1 MeV. 
Due to the very low level of background, states with an $s_\Lambda$ coupled 
to excited $^{11}$B core states are clearly observed between the g.s. and the 
level at 10.93 MeV  with signal to noise ratios  larger than 5. The 
positions of these levels was determined with uncertainties less than 100 keV. 
Cross sections are determined at the level of 15--20\%. As in the
Hall C experiments~\cite{tang14}, a measurable strength with good 
energy resolution has been observed in the core-excited part of 
the spectrum. This is due the fact that the spin-spin interaction enhances 
the cross sections for these states with respect to the weak-coupling limit
(compare the structure factors on the right in Fig.~\ref{lb12} with $C^2S$
on the left).

\subsubsection{The $^{16}$O target}
%
\begin{figure}
\centering
\includegraphics[width=6.2cm,angle=270]{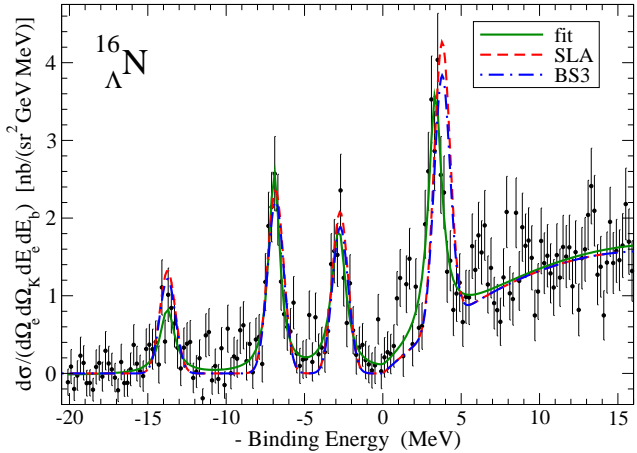}
\caption{The \lam{16}{N} binding-energy spectrum. The solid line shows
the best fit using Voigt functions (see text for details). The theoretical 
curves (dashed and dash-dotted liness) were calculated with an average width
extracted from the fit (FWHM = 1177 keV).}
\label{oxy3}
\end{figure}

%
\begin{table*}[t]
\caption{Excitation energies, widths, and cross sections obtained by fitting 
the $^{16}$O$(e,e'K^+)$\lam{16}{N} spectrum (first three columns) compared with 
theoretical predictions using the SLA and BS3 models for the elementary
interaction (next six columns). 
\label{tab:Oresults}}
\begin{ruledtabular}
\begin{tabular}{ccccccccc}
\multicolumn{3}{c}{Experimental data}  &
\multicolumn{6}{c}{Theoretical cross sections for the SLA and BS3 models} \\
$E_x$ &  Width     &  Cross section  & $E_x$ & $J^\pi$ & 
\multicolumn{2}{c}{SLA} & \multicolumn{2}{c}{BS3} \\
(MeV) & (FWHM, MeV)& $(nb/sr^2/GeV)$ & (MeV) &         & 
$(nb/sr^2/GeV)$  & Sum  & $(nb/sr^2/GeV)$  &  Sum \\
 \hline
0.00 $\pm$ 0.02  &  1.71 $\pm$ 0.70  &  1.45 $\pm$ 0.26 & 
         0.000   &  $0^-$  &  0.003  &       &  0.134 &       \\
 & & &   0.023   &  $1^-$  &  1.657  &  1.66 &  1.391 &  1.52 \\
 & & & & & & & &\\
6.83 $\pm$ 0.06  &  0.88 $\pm$ 0.31  &  3.16 $\pm$ 0.35 & 
         6.730   &  $1^-$  &  0.818  &       &  0.688 &       \\
 & & &   6.978   &  $2^-$  &  2.201  & 3.02  &  2.153 &  2.84 \\
 & & & & & & & &\\
10.92 $\pm$ 0.07 &  0.99 $\pm$ 0.29  &  2.11 $\pm$ 0.37 & 
         11.000  &  $2^+$  &  1.948  &       &  1.627 &        \\
 & & &   11.116  &  $1^+$  &  0.607  &       &  0.679 &        \\
 & & &   11.249  &  $1^+$  &  0.069  & 2.62  &  0.071 &  2.38  \\
 & & & & & & & &\\
17.10 $\pm$ 0.07 &  1.00 $\pm$ 0.23  &  3.44 $\pm$ 0.52 & 
         17.303  &  $1^+$  &  0.166  &       &  0.181 &        \\
 & & &   17.515  &  $3^+$  &  2.311  &       &  2.045 &    \\
 & & &   17.567  &  $2^+$  &  2.071  &  4.55  &  1.723 & 3.95  \\
\end{tabular}
\end{ruledtabular}
\end{table*}%
%
\begin{table*}
\caption{Central kinematics of the Hall C experiments with the carbon 
target in the laboratory reference frame. 
The values of $E_i$, $E_f$, $\theta_e$, 
$\theta_{Ke}$, and $\Phi_K= 90^{o}$ were used in our calculations. 
\label{tab:HallC_kinematics}}
\begin{ruledtabular}
\begin{tabular}{ccccccccccc} 
Experiment & $E_i$  & $E_f$ & $\theta_e$ & $\theta_{Ke}$ & $E_\gamma$ & 
$\theta_{\gamma e}$ & $Q^{2}$ & $\epsilon$ & $\Gamma$ &   $p_K$   \\
       & [GeV]  & [GeV] &   [deg]    &   [deg]       &  [GeV]     &
   [deg]            & [GeV$^2$] &      &  [(GeV sr)$^{-1}$]& [GeV]\\     
\hline
E01-011 & 1.851 & 0.351  &  5.4        &   7.11       &   1.50 & 
    1.26             & 0.00577   &  0.365     & 0.0287   &  1.20--1.22 \\
E05-115 & 2.344 & 0.844  &  5.4        &   7.62       &   1.50 &
    3.02             & 0.01756   &  0.635     & 0.0310   &  1.20--1.22 \\
\end{tabular}
\end{ruledtabular}
\end{table*}

$^{16}$O targets have been extensively used in hypernuclear
studies with the $(K^-,\pi^-)$, $(\pi^+, K^+)$, and
$(K^-_\mathrm{stop},\pi^-)$ reactions with dominant non-spin-flip
reaction mechanisms that excite natural-parity states~\cite{hashtam06}. 
In all cases, four peaks are seen with the excited states at
$\approx 6.2$, $\approx 10.6$, and $\approx 17.1$ MeV corresponding
to $\Lambda$'s in $s$ and $p$ orbits coupled to the $p^{-1}_{1/2}$ 
ground state and the 6.176-MeV $p^{-1}_{3/2}$ state of $^{15}$O.
In the simple particle-hole limit, the degenerate multiplets contain 
2, 2, 4, and 6 states, respectively, and the cross sections would
be in the ratio 2:1 for peaks based on the $p_{3/2}$ vs. $p_{1/2}$
hole states. The first two peaks correspond to $1^-$ states
and the $B_\Lambda$ value for the lowest $1^-$ state is not 
particularly well determined. In the CERN $(K^-,\pi^-)$ 
experiment~\cite{brueckner78}, the third and fourth peaks correspond to 
substitutional $0^+$ states. At the larger momentum transfer of the 
stopped $K^-$ work at KEK~\cite{tamura94}, the same peaks contain 
contributions from both $0^+$ and $2^+$ hypernuclear states.  
In the $(\pi^+, K^+)$ reaction, first performed at BNL~\cite{pile91} 
and later at KEK~\cite{hashtam06} with better energy 
resolution, only the $2^+$ states are expected to contribute.
Finally from $\gamma$-ray spectroscopy, the $0^-$ state in \lam{16}{O}
is the ground state, the ground-state doublet spacing is 26.4 keV,
and the $1^-$ and $2^-$ states of the excited doublet are at 6562 and 
6786 keV, respectively~\cite{ukai08}.

The experimental knowledge can be enhanced using the $(e,e'K^+)$
electroproduction reaction characterized  by a large momentum transfer
to the hypernucleus ($q \gtrsim$ 250 MeV/c) and strong spin-flip terms,
even at zero degree $K^+$ production angles, resulting in the excitation
of both natural- and unnatural-parity states. In the present case,
$1^-$, $2^-$, $1^+$, $2^+$, and $3^+$ particle-hole states can be
excited with significant cross sections. In addition, the
$K^+\Lambda$ associated production occurs on a proton making
\lam{16}{N}, the mirror to \lam{16}{O}. After taking into account 
that the $p_{3/2}$-hole state is 148 keV higher in $^{15}$N than 
$^{15}$O, comparison of the energy spectra (and especially of 
$\Lambda$ binding energies) of these mirror hypernuclei can, 
in principle, shed light on charge-dependent effects in 
hyperon-nucleon interactions. 

The binding-energy spectrum for \lam{16}{N} electroproduction is shown in 
Fig.~\ref{oxy3} where the experimental data from this experiment (the points 
with statistical errors)~\cite{cusanno09} are compared with theoretical 
predictions (dashed and dash-dotted lines). The fit to the data (solid line)
 has been made using Voigt functions, which in our case were the convolution 
of a narrow Gaussian with FWHM= 330 keV and the Breit-Wigner form of 
independent widths. Practically a zero constant background up to the 
quasi-free threshold at 13.76 MeV and $\chi_{n.d.f.}^2$ = 1.01 were obtained. 
In the quasifree region, a quadratic form of background was assumed, similar 
to the case of \lam{12}{B}. The FWHM around 1000 keV was obtained for all 
peaks consistent with the \lam{12}{B} case. 

\begin{table*}[t]
\caption{Comparison of theoretical predictions using the SLA and BS3 models
for the elementary cross section for the two-folded cross 
sections in $^{12}$C$(e,e'K^+)$\lam{12}{B} with data from the Hall C 
experiments~\cite{tang14}. The excitation energy was determined to 
be $11.517 - B_\Lambda$ MeV in E01-011 and $11.529 - B_\Lambda$ MeV in E05-115
where $B_\Lambda$ is the $\Lambda$ binding energy. The 8th and 10th columns 
give the summed cross section for assigned multiplets, to 
compare with the experimental result in the fourth column.
\label{tab:HallC_Cresults}}
\begin{ruledtabular}
\begin{tabular}{cccccccccc}
\multicolumn{10}{c}{Experiment E01-011} \\
\multicolumn{4}{c}{Experimental data}  &
\multicolumn{6}{c}{Theoretical cross sections for the SLA and BS3 models} \\
Peak & $B_\Lambda$ & $E_x$ & Cross section & $E_x$ & $J^\pi$ & 
\multicolumn{2}{c}{SLA} &\multicolumn{2}{c}{BS3}\\
 No. &  (MeV)      & (MeV) &  $(nb/sr)$    & (MeV) &  & $(nb/sr)$ & Sum & $(nb/sr)$ & Sum\\
 \hline
 1 &11.517 $\pm$ 0.031 &  0.0   & 101.0 $\pm$ 4.2 &  0.0   & $1^-$ & 13.90 &       & 14.04 &       \\
   &                   &        &                 &  0.116 & $2^-$ & 44.70 & 58.60 & 35.33 & 49.37 \\
 2 & 8.390 $\pm$ 0.075 &  3.127 &  33.5 $\pm$11.3 &  2.587 & $1^-$ & 17.26 &       & 14.65 &       \\
   &                   &        &                 &  2.593 & $0^-$ &  0.04 & 17.31 &  0.12 & 14.76 \\
 3 & 5.440 $\pm$ 0.085 &  6.077 &  26.0 $\pm$ 8.8 &  4.761 & $2^-$ &  0.37 &       &  0.30 &       \\
   &                   &        &                 &  5.642 & $2^-$ &  7.20 &       &  5.69 &       \\ 
   &                   &        &                 &  5.717 & $1^-$ &  2.44 & 10.01 &  2.24 &  8.22 \\
 4 & 2.882 $\pm$ 0.085 &  8.635 &  20.5 $\pm$ 7.3 &        &       &       &       &       &       \\
 5 & 1.470 $\pm$ 0.091 & 10.047 &  31.5 $\pm$ 7.4 & 10.480 & $2^+$ &  5.15 &       &  5.16 &       \\
   &                   &        &                 & 10.525 & $1^+$ &  2.16 &  7.31 &  1.77 &  6.93 \\
 6 & 0.548 $\pm$ 0.035 & 10.969 &  87.7 $\pm$15.4 & 11.059 & $2^+$ & 25.23 &       & 22.35 &       \\ 
   &                   &        &                 & 11.132 & $3^+$ & 39.08 &       & 29.70 &       \\
   &                   &        &                 & 11.197 & $0^+$ &  0.10 & 64.41 &  0.42 & 52.47 \\
 7 &-0.318 $\pm$ 0.085 & 11.835 &  46.3 $\pm$10.3 & 11.674 & $1^+$ &  5.37 &  5.37 &  4.25 &  4.25 \\
 8 &-0.849 $\pm$ 0.101 & 12.366 &  28.5 $\pm$ 7.4 & 12.967 & $2^+$ & 13.96 &       & 12.37 &       \\
   &                   &        &                 & 13.074 & $1^+$ &  4.36 & 18.32 &  3.57 & 15.93 \\
\hline \\ 
\multicolumn{10}{c}{Experiment E05-115} \\
\multicolumn{4}{c}{Experimental data}  &
\multicolumn{6}{c}{Theoretical cross sections for the SLA and BS3 models} \\
Peak & $B_\Lambda$ & $E_x$ & Cross section & $E_x$ & $J^\pi$ & 
\multicolumn{2}{c}{SLA} &\multicolumn{2}{c}{BS3}\\
 No. &  (MeV)      & (MeV) &  $(nb/sr)$    & (MeV) &  & $(nb/sr)$ & Sum & $(nb/sr)$ & Sum\\
 \hline 
 1 &11.529 $\pm$ 0.025 &  0.0   &  83.0 $\pm$ 3.0 &  0.0   & $1^-$ & 13.14 &       & 13.53 &       \\
   &                   &        &                 &  0.116 & $2^-$ & 42.05 & 55.19 & 33.49 & 47.02 \\
 2 & 8.425 $\pm$ 0.047 &  3.104 &  19.1 $\pm$ 3.7 &  2.587 & $1^-$ & 16.24 &       & 13.62 &       \\
   &                   &        &                 &  2.593 & $0^-$ &  0.07 & 16.31 &  0.35 & 13.98 \\
 3 & 5.488 $\pm$ 0.052 &  6.041 &  18.0 $\pm$ 4.6 &  4.761 & $2^-$ &  0.35 &       &  0.28 &       \\
   &                   &        &                 &  5.642 & $2^-$ &  6.76 &       &  5.39 &       \\
   &                   &        &                 &  5.717 & $1^-$ &  2.29 &  9.40 &  2.11 &  7.78 \\
 4 & 2.499 $\pm$ 0.075 &  9.030 &  16.2 $\pm$ 5.1 &        &       &       &       &       &       \\
 5 & 1.220 $\pm$ 0.056 & 10.309 &  28.7 $\pm$ 7.2 & 10.480 & $2^+$ &  4.90 &       &  4.98 &       \\
   &                   &        &                 & 10.525 & $1^+$ &  2.07 &  6.97 &  1.78 &  6.77 \\
 6 & 0.524 $\pm$ 0.024 & 11.005 &  75.7 $\pm$10.8 & 11.059 & $2^+$ & 23.97 &       & 21.24 &       \\
   &                   &        &                 & 11.132 & $3^+$ & 41.17 &       & 32.40 &       \\
   &                   &        &                 & 11.197 & $0^+$ &  0.12 & 65.26 &  0.49 & 54.12 \\
 7 &-0.223 $\pm$ 0.039 & 11.752 &  39.0 $\pm$ 7.4 & 11.674 & $1^+$ &  5.45 &  5.45 &  4.32 &  4.32 \\
 8 &-1.047 $\pm$ 0.078 & 12.576 &  27.8 $\pm$ 7.9 & 12.967 & $2^+$ & 13.25 &       & 11.69 &       \\
   &                   &        &                 & 13.074 & $1^+$ &  4.18 & 17.44 &  3.57 & 15.27 \\    
\end{tabular}
\end{ruledtabular}
\end{table*}

The values of excitation energies, widths, and cross sections extracted from 
the fit are given in Table~\ref{tab:Oresults} together with the predicted cross 
sections for the lowest states of \lam{16}{N} (the structure comes from a 
simple particle-hole calculation). Only statistical errors are 
reported for the measured cross sections. Systematic errors, dominated by 
uncertainty in the target thickness, are at the $20\%$ level. In the DWIA 
calculations, improved kaon distortion and better Woods-Saxon radial wave 
functions were used in comparison with our previous 
calculations~\cite{cusanno09}. 
In the old calculations only the proton in $p_{1/2}$ state and also 
a stronger kaon absorption were considered where the latter had reduced the 
cross sections. In the new computation of the Woods-Saxon wave functions 
the target proton was bound by 17.82 and 11.20 MeV for the $p_{3/2}$ and 
$p_{1/2}$ states, respectively and the final $\Lambda$ was considered to be 
bound by 13.5, 2.3, and 2.9 MeV for the $s_{1/2}$, $p_{1/2}$, and $p_{3/2}$ 
states, respectively.

Four peaks are observed in the spectrum. The ground-state peak gives a 
$\Lambda$ separation energy of $B_\Lambda\!=\!13.76\pm 0.16$ MeV for the $1^-$ 
member of the ground-state doublet in \lam{16}{N}. Three more peaks are 
observed at binding energies of $6.93$, $2.84$, and $-3.34$ MeV. 
The theory overpredicts the cross sections by 10--30\% in contrary to 
the case of \lam{12}{B} and \lamb{9}{Li} production. This opposite tendency 
of the hypernuclear cross sections can be hardly attributed to uncertainty 
in the elementary production cross sections~\cite{bydzovsky12,bydzovsky07}. 
The overpredicted cross section is more likely due the use of simple hole
states for the $^{15}$N core nucleus because the analysis of the 
$^{16}$O$(e,e'p)^{15}$N reaction shows that the spectroscopic factors for 
proton removal are reduced from their simple shell-model values and that the 
discrete $3/2^-$ strength is spread over four states~\cite{leuschner94}, a 
fact that has to be explained by a multi-$\hbar\omega$ shell-model 
calculation. This argument does not work to explain the underpredicted cross 
section for the $^{12}$C target because a similar analysis of the $(e,e'p)$ 
reaction in this case again shows the usual reduction in spectroscopic 
factors with respect to $p$-shell values~\cite{vandersteenhoven88}. 

\subsection{Hall C data}
\label{comparison_AC}

 To compare our theoretical results for \lam{12}{B} production with data 
from the Hall C experiments E01-011 and E05-115~\cite{tang14}, we use
the kinematics presented in Table~\ref{tab:HallC_kinematics} for the values 
of $E_i$, $E_f$, $\theta_e$, $\theta_{Ke}$, and $\Phi_K= 90^{o}$. 
Note that the virtual-photon energy and mass ($Q^2$) and the kaon momentum 
significantly differ in the Hall A and C measurements, see 
Tables~\ref{tab:kinematics} and \ref{tab:HallC_kinematics}. 
  
The results are presented in Table ~\ref{tab:HallC_Cresults}.
As in Table~\ref{tab:Cresults} for the Hall A experiment, the theoretical cross
sections are  30--50\% smaller than the experimental values suggesting that 
this phenomenon is present in a broader beam-energy region.
In Table~\ref{tab:HallC_Cresults}, we make assignments with certainty 
only for the $\Lambda$ s-wave states and leave open an assignment for the 
higher states. 

Note that especially for E01-011 kinematics with very small $Q^2$, given in
Table~\ref{tab:HallC_kinematics}, the cross section is dominated by the 
transverse contributions and therefore photoproduction 
calculations~\cite{bydzovsky12,motoba15} are justified.
Here, for the photon lab energy 1.5 GeV, the SLA model gives larger
elementary cross sections than BS3 (see Fig.5 in Ref.~\cite{skoupil18}) 
and therefore the hypernucleus cross sections are again larger for SLA 
than for BS3 as in the case of Hall A kinematics. We recall that in the 
Hall A case, smaller predictions of BS3 were due to steeper descent of the
transverse elementary cross section with $Q^2$.

 Finally, the Hall C fit to their data included five peaks in the region
defined by $\sim 2$ MeV on either side of the dominant $p_\Lambda$ peak.
Such peaks are not unexpected because states with an $s_\Lambda$ coupled
to the $3/2^+$ and $5/2^+$ core states shown at the left of Fig.~\ref{lb12}
exist and can mix with the $p^7p_\Lambda$ states and aquire some formation
strength. These two types of states are both $1\hbar\omega$ states and
one must eliminate spurious linear combinations from the full $1\hbar\omega$
shell-model basis~\cite{motoba15} and this, by itself, enforces mixtures
of the $s_\Lambda$ and $p_\Lambda$ states.

\section{Summary and conclusions}

The systematic study of hypernuclear spectroscopy by electroproduction of 
strangeness performed at Jefferson Laboratory in Hall A has been very 
successful. It has provided important elements for a better understanding of 
the baryon-baryon interactions and production mechanism in strangeness physics.
The experiment was successful but challenging because important modifications 
to the Hall A apparatus were needed. The new experimental equipment, aerogel 
threshold detectors, septum magnets, and the RICH detector all gave excellent 
performance. Sub-MeV energy resolution and very clean, background free, 
spectra were obtained. The results of the hypernuclear spectroscopy performed
on $^{12}$C, $^{16}$O, and $^{9}$Be targets provide important data for a 
better understanding of strangeness physics.  Results from $^{12}$C showed  
significant strength in the core-excited part of the spectrum. 
The spectrum is quite well reproduced by the theory apart from an overall 
underestimation of the experimental cross section.
Moreover, for $^{16}$O, thanks to the calibration 
with the hydrogen present in the waterfall target, a very precise 
determination of the $\Lambda$ binding energy for \lam{16}{N} was obtained. 
In the case of $^{9}Be$ the measured cross sections are in good agreement for 
the first peak with the values predicted using the SLA model and simple 
shell-model wave function. The reason for the disagreement in strength for 
the second and third peak is hard to ascertain and could be due to a number
of deficiences in the structure or reaction calculations.

 We now list the improvements that have been made with respect to our
previous separate publications on the data from the $^{12}$C~\cite{iodice07},
$^{16}$O~\cite{cusanno09}, and $^9$Be~\cite{urciuoli15} targets.

\begin{itemize}

\item A new data point on the elementary electroproduction reaction at the 
forward-angle kinematics of the E94-107 experiment is presented. Given
the lack of electroproduction data at the forward angles important for
hypernuclear electroproduction, this is an important measurement. A detailed 
comparison of existing data on the elementary reaction (mostly photoproduction)
with a wide range of models is presented. 

\item For the hypernuclear electroproduction, results from the BS3
isobar model are given for comparison with the SLA model used previously.

\item A new analysis of the carbon data was made in which radiative corrections 
were performed as for the Be target. The spectrum was improved and the
extracted peak widths are more consistent now.

\item New structure and reaction calculations for carbon have been made that 
use the complete set of $p$-shell core states in the structure calculations, 
corrected kaon distortion, include hypernuclear recoil, and use  
realistic Woods-Saxon wave functions.

\item The theory (consistent calculations) was compared with data for several 
 targets  (Be, C, O) and  for the carbon target for the different kinematics of 
Hall A and C experiments. This implies a test of the reaction mechanism for 
DWIA calculations.

\item New calculations were made for the beryllium target. We used a new 
structure (fit4), the improved kaon distortion, the Woods-Saxon wave 
functions, and included the hypernuclear recoil.

\item Slightly improved calculations were made for the oxygen target, using
improved kaon distortion, and elaborated using the Woods-Saon wave functions.

\end{itemize}

In conclusion, we can also say that a more detailed analysis of the DWIA 
calculations using different elementary production amplitudes and
larger-basis shell-model calculations would be interesting  
(mainly due to our findings on the opposite discrepancies in
cross sections for the $^{12}$C and $^{16}$O targets).

We acknowledge the Jefferson Lab Physics and Accelerator Division staff for 
their outstanding efforts that made this work possible. We would like to
single out the contributions made by our deceased colleagues Salvatore 
Frullani (significant contributions from the beginning of this project), 
Francesco Cusanno (leader of the analysis for the oxygen target), and
Miloslav Sotona (reaction calculations for the three published papers). 
This work was supported 
by U.S. DOE contract DE-AC05-84ER40150, Mod. nr. 175, under which the 
Southeastern Universities Research Association (SURA) operates the Thomas 
Jefferson National Accelerator Facility, the Italian Istituto Nazionale di 
Fisica Nucleare and by the Grant Agency of the Czech Republic under grant No.
P203/15/04301, by the French CEA and CNRS/IN2P3, and by the U.S. DOE 
under contracts, DE-AC02-06CH11357, DE-FG02-99ER41110, and DE-AC02-98-CH10886, 
and by the U.S. National Science Foundation.

\bibliography{archival}

\end{document}